%% file: ms_DECaLS_SL.tex
\PassOptionsToPackage{dvipsnames}{xcolor}
\documentclass[preprint]{aastex62-no-date}

\usepackage{amsmath}
\usepackage{capt-of}   
\usepackage{empheq}
\usepackage{enumitem}
\usepackage{hhline}
\usepackage{longtable}
\usepackage{numprint}
\usepackage{pifont}
\usepackage{xspace}
\usepackage{placeins}
\usepackage{graphicx}
\usepackage{mwe}

\newcommand{\rf}{\textcolor{black}}
\newcommand{\rff}{\textcolor{black}}
\newcommand{\ed}{\textcolor{black}}

\newcommand{\HST}{\emph{HST}\xspace}

\newcommand{\lenstot}{342\xspace}
\newcommand{\lenstotnew}{335\xspace}
\newcommand{\lensz}{121\xspace}

\newcommand{\lensA}{60\xspace}

\newcommand{\lensAz}{Thirty~three\xspace}
\newcommand{\lensAnew}{53\xspace}
\newcommand{\lensB}{106\xspace}
\newcommand{\lensBnew}{106\xspace}

\newcommand{\lensBz}{Thirty~seven\xspace}
\newcommand{\lensC}{176\xspace}
\newcommand{\lensCnew}{176\xspace}
\newcommand{\zerrmaxC}{3.9\xspace}
\newcommand{\lensCz}{Fifty~one\xspace}

\newcommand{\twopr}{^{\prime \prime}}

\accepted{March 12, 2020}
\submitjournal{ApJ}

\shorttitle{DESI DECaLS Strong Lenses}
\shortauthors{Huang, Domingo, Pilon, Ravi, Storfer, Schlegel et al.}

\begin{document}
\title{Finding Strong Gravitational Lenses in the DESI DECam Legacy Survey}


\author[0000-0001-8156-0330]{X.~Huang}
\affiliation{Department of Physics and Astronomy, University of San Francisco, San Francisco, CA 94117-1080}

\author{C.~Storfer}
\affiliation{Department of Physics and Astronomy, University of San Francisco, San Francisco, CA 94117-1080}

\author{V.~Ravi}
\affiliation{Department of Computer Science, University of San Francisco, San Francisco, CA 94117-1080}

\author{A.~Pilon}
\affiliation{Department of Physics and Astronomy, University of San Francisco, San Francisco, CA 94117-1080}

\author{M.~Domingo}
\affiliation{Department of Computer Science, University of San Francisco, San Francisco, CA 94117-1080}

\author[0000-0002-5042-5088]{D.J.~Schlegel}
\affiliation{Physics Division, Lawrence Berkeley National Laboratory, 1 Cyclotron Road, Berkeley, CA, 94720}


\author[0000-0002-5042-5088]{S.~Bailey}
\affiliation{Physics Division, Lawrence Berkeley National Laboratory, 1 Cyclotron Road, Berkeley, CA, 94720}

\author[0000-0002-4928-4003]{A.~Dey}
\affiliation{National Optical Astronomy Observatory, 950 N. Cherry Ave., Tucson, AZ 85719}

\author[0000-0002-5042-5088]{R.R.~Gupta}
\affiliation{Physics Division, Lawrence Berkeley National Laboratory, 1 Cyclotron Road, Berkeley, CA, 94720}

\author[0000-0003-2092-6727]{D.~Herrera}
\affiliation{National Optical Astronomy Observatory, 950 N. Cherry Ave., Tucson, AZ 85719}

\author[0000-0002-0000-2394]{S.~Juneau}
\affiliation{National Optical Astronomy Observatory, 950 N. Cherry Ave., Tucson, AZ 85719}

\author[0000-0003-1838-8528]{M.~Landriau}
\affiliation{Physics Division, Lawrence Berkeley National Laboratory, 1 Cyclotron Road, Berkeley, CA, 94720}

\author[0000-0002-1172-0754]{D.~Lang}
\affiliation{Dunlap Institute, University of Toronto, Toronto, ON M5S 3H4, Canada}
\affiliation{Department of Astronomy \& Astrophysics, University of Toronto, Toronto, ON M5S 3H4, Canada} 
\affiliation{Perimeter Institute for Theoretical Physics, Waterloo, ON N2L 2Y5, Canada}

\author[0000-0002-1125-7384]{A.~Meisner}
\affiliation{National Optical Astronomy Observatory, 950 N. Cherry Ave., Tucson, AZ 85719}

\author[0000-0002-2733-4559]{J.~Moustakas}
\affiliation{Department of Physics and Astronomy, Siena College, 515 Loudon Rd., Loudonville, NY 12211}

\author{\ed{A.D.~Myers}}
\affiliation{Department of Physics \& Astronomy, University of Wyoming, 1000 E. University, Dept 3905, Laramie, WY 8207}

\author[0000-0002-3569-7421]{E.F.~Schlafly}
\affiliation{National Optical Astronomy Observatory, 950 N. Cherry Ave., Tucson, AZ 85719}

\author[0000-0001-5567-1301]{F.~Valdes}
\affiliation{National Optical Astronomy Observatory, 950 N. Cherry Ave., Tucson, AZ 85719}

\author{B.A.~Weaver}
\affiliation{National Optical Astronomy Observatory, 950 N. Cherry Ave., Tucson, AZ 85719}

\author[0000-0001-5287-4242]{J.~Yang}
\affiliation{Steward Observatory, University of Arizona, 933 N. Cherry Ave., Tucson, AZ 85721}

\author{C.~Y\`eche}
\affiliation{IRFU, CEA, Universit\'e Paris-Saclay, F-91191 Gif-sur-Yvette, France}


\begin{abstract}
\input{abstr}

\end{abstract}

\section{Introduction}
\label{sec:intro}
\input{intro}

\FloatBarrier
\section{Observations}
\label{sec:observations}
\input{observations}

\section{The Training Sample and Residual Neural Networks}
\label{sec:resnet}
\input{model-training}

\section{Results}
\label{sec:results}
\input{results}
\section{Discussion}\label{sec:discussion}

\input{discussion}

\section{Conclusions}\label{sec:conclusion}
\input{conclusion}
\section{Acknowledgement}\label{sec:acknowledgement}
\input{acknowledgement}

\bibliographystyle{aasjournal}
\bibliography{dustarchive}

\end{document}

%% file: abstr.tex
We perform a semi-automated search for strong gravitational lensing systems in the 9,000~deg$^2$ Dark Energy Camera Legacy Survey (DECaLS), \ed{part of the DESI Legacy Imaging Surveys (Dey et al.).}
The combination of the depth and breadth of these surveys are unparalleled at this time,  
making them particularly suitable for discovering new strong gravitational lensing systems.  
We adopt the deep residual neural network architecture (He et al.) developed by Lanusse et al. for the purpose of finding strong lenses in photometric surveys.
We compile a training sample that consists of  
known lensing systems 
in the Legacy Surveys and the Dark Energy Survey as well as  
non-lenses in the footprint of DECaLS.
In this paper we show the results of applying our trained neural network to the cutout images centered on galaxies typed as ellipticals (Lang et al.) in DECaLS.
The images that receive the highest scores (probabilities)
are \ed{visually inspected and ranked}.
Here we present \ed{\lenstotnew} candidate strong lensing systems, identified for the first time.

%% file: intro.tex
Strong lensing systems \ed{\citep{walsh1979a, lynds1986a, soucail1987a, soucail1988a, paczynski1987a}} have been used to study how dark matter is distributed in galaxies and clusters 
\citep[e.g.,][]{kochanek1991a, koopmans2002a, bolton2006a, koopmans2006a, vegetti2009a, tessore2016a}. 
 As a cosmological probe, time delays in multiply lensed quasars provide competitive constraints on the Hubble constant $H_\mathrm{0}$ 
\citep[e.g.][]{refsdal1964a, blandford1992a, suyu2010a, suyu2013a, treu2016a, bonvin2017a, wong2019a} independent of the distance ladder approach.

In recent years, highly magnified, multiply imaged supernovae (SNe), both core-collapse \citep{kelly2015a} and Type Ia \citep{quimby2014a, goobar2017a}, have been discovered.
With their well-characterized brightness time evolution in optical and near-infrared wavelengths (the SN lightcurves), 
such strongly lensed SNe are ideally suited to measure time-delays and $H_\mathrm{0}$
in future surveys \citep[e.g.,][]{goldstein2017a, goldstein2018a, goldstein2018b, wojtak2019a} .  

In this paper we show that 
hundreds of new strong lensing systems can be found in the three band imaging data ($grz$) from the Dark Energy Spectroscopic Instrument (DESI) Legacy Surveys\footnote{\url{legacysurvey.org}}
 \citep{dey2019a}.
 To find these lenses from a data set that covers one third of the sky, we adopt the residual neural network \citep{he2015a, he2015b, he2016a} developed by \citet{lanusse2018a}, 
the winning algorithm of the Strong Gravitational Lens Finding Challenge \citep{metcalf2018a}, to automate the process as much as possible.

This paper is organized as follows.  
A brief description of the Legacy Surveys is given in \S~\ref{sec:observations}.  
In \S~\ref{sec:resnet}, we describe our methodology and training sample.
In \S~\ref{sec:results}, we show the inference results and present our best strong lensing system candidates.  
We discuss our results in \S~\ref{sec:discussion}, and conclude in \S~\ref{sec:conclusion}.



%% file: observations.tex
The details of the DESI Legacy Imaging Surveys are described in \citet[][D19]{dey2019a}.  Here we present a brief summary.
The Legacy Surveys consist of three projects: the Dark Energy Camera Legacy Survey (DECaLS), observed by the  Dark Energy Camera \citep[DECam;][]{flaugher2015a} on the 4-m Blanco telescope at the Cerro Tololo Inter-American Observatory; 
the Beijing-Arizona Sky Survey (BASS), by the 90Prime camera \ed{\citep{williams2004a}} on the Bok 2.3-m telescope owned and operated by the University of Arizona located on Kitt Peak;
and the Mayall $z$-band Legacy Survey (MzLS), by the Mosaic3 camera \ed{\citep{dey2016a}} on the 4-meter Mayall telescope at Kitt Peak National Observatory. 
\ed{Together they will ultimately cover}
$\sim$14,000 deg$^2$ of the extragalactic sky visible from the northern hemisphere in $grz$ bands,
with a $5\,\sigma$ $z$-band median limiting magnitude of 22.5~mag for galaxies with an exponential disk profile with $\mathrm{r}_{i, \, \mathrm{half}} = 0.45\twopr$. 

The combined survey footprint is split into two contiguous areas by the Galactic plane. 
DECaLS covers the $\sim 9000$ deg$^2$ $\delta \lesssim +32^\circ$ sub-region of the Legacy Surveys. 
\rf{Figure~\ref{fig:decals} shows the different regions in the the Legacy Surveys footprint and the depth of the $z$-band observation for DR7, which is indicative of the completeness of the survey.}  

\begin{minipage}{\linewidth}
\makebox[\linewidth]{
  \includegraphics[keepaspectratio=true,scale=0.48]{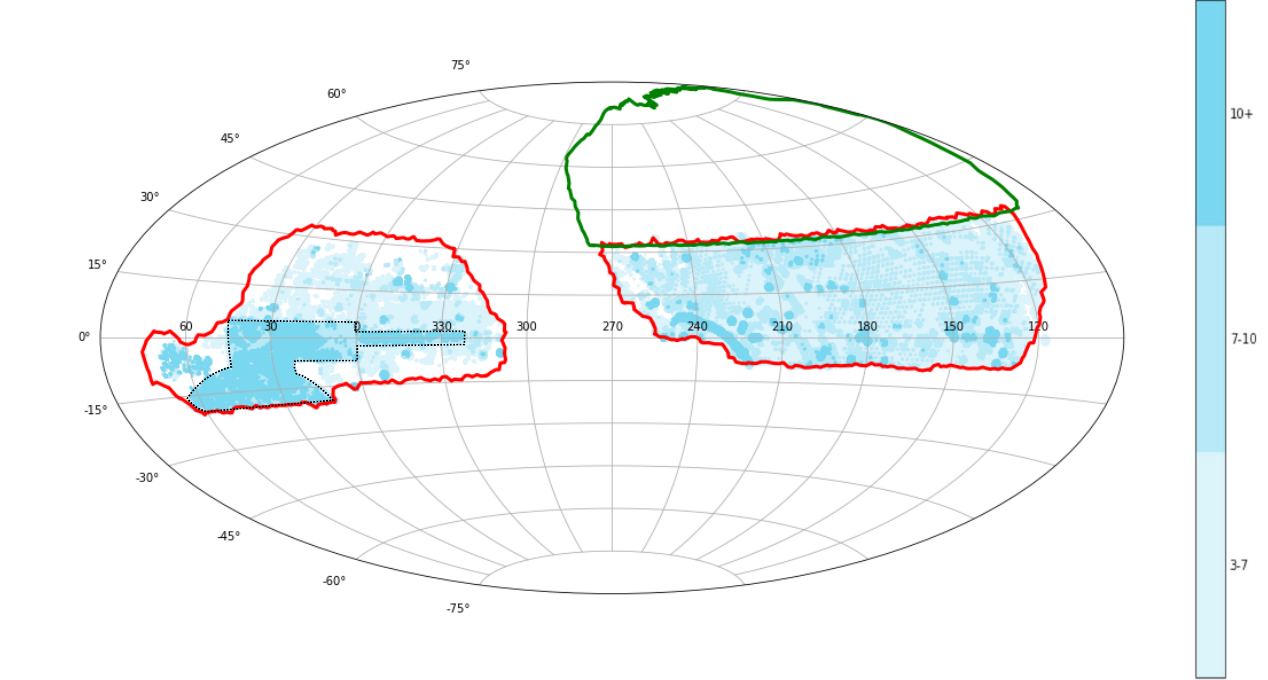}}
\captionof{figure}{\rf{
The DESI Legacy Surveys footprint in an equal area Aitoff projection in equatorial coordinates with the DECaLS and MzLS/BASS regions highlighted with red and green outlines, respectively.
Above the Galactic plane, 
there is a small amount of overlap between the surveys.
For DECaLS, patches with different shades of blue indicate the depth in $z$ band: 
light blue for between three and seven passes;
medium blue, between seven and ten; 
and dark blue, greater than ten. 
We left the MzLS/BASS region blank because in the paper we have not searched for lenses in that region.
Note that observations in $grz$ bands of the northern part (Dec$> -30^\circ$) of the Dark Energy Survey (DES; black dotted outline) is included in DECaLS. 
}}
\label{fig:decals}
\end{minipage}

\rf{For DECaLS, the delivered image quality has FWHM of approximately 1.29, 1.18, $1.11\twopr$ for $g$, $r$, and $z$ bands respectively.}  
The MzLS has imaged the $\delta \gtrsim +32^\circ$ (NGC) footprint of the Legacy Surveys in $z$-band that complemented the BASS $g$- and $r$-band observations in the same \ed{sub-region}.
While the delivered image quality of MzLS has a median seeing of $\approx 1.01 \twopr$, the median FWHM's 
for BASS are \rf{$1.61\twopr$} and \rf{$1.47\twopr$} in the $g$- and $r$-bands, respectively. 
\ed{We choose here to focus on DECaLS due to its better $gr$ seeing than BASS.  
However, we intend in future work to apply the machine-learning framework we have developed for DECaLS to the northern BASS/MzLS area,} 
\rf{by possibly incorportating transfer learning, active learning, or domain adaption \citep[e.g.,][]{tzeng2017a} techniques.} 

The Legacy Surveys used the Tractor package \citep{lang2016a} 
as a forward-modeling approach to perform source extraction on pixel-level data. 
Tractor takes as input the individual images from multiple exposures in multiple bands, with different seeing in each. 
After source detection,
the point source (``PSF") and spatially extended (``REX", round exponential galaxy) models are computed for every source and the better of these two is used when deciding whether to keep the source.
The spatially extended sources (REX) are further classified if $\chi^2$ is improved by 9 
by treating it as a deVaucouleurs (DEV), an exponential (EXP) profile, or a composite of deVaucouleurs + exponential (COMP)\footnote{\url{http://legacysurvey.org/dr7/description/}}.  
The same light profile (EXP, DEV, or COMP) is consistently fit to all images in order to determine the best-fit source shape parameters and photometry.


The categories of DEV and COMP 
indicate the classification of elliptical galaxies.  
Given that the vast majority of lensing events are caused by early type galaxies, 
we decided to target only objects with DEV and COMP classifications in this paper. 


%% file: model-training.tex
\subsection{Training Sample}\label{sec:train}

Deep convolutional neural networks (CNNs) 
and their variations have been shown to be highly effective in image recognition.  
In recent years, this technique has been successfully applied to recognize instances of strong lenses in simulations \citep[e.g.,][and references therein]{metcalf2018a}.
\ed{In previous applications of CNNs to real observations, training samples are constructed from simulated lens images, 
combined with observed \citep{petrillo2017a}, 
simulated \citep{pourrahmani2018a, jacobs2017a}, 
or a mixture of observed and simulated non-lenses  \citep{jacobs2019a}.}
This is because the number of known lenses, on the order of several hundred, is thought to be too small to effectively train CNN models.  
We note that the data set for \ed{\citet{jacobs2017a} is from the Canada-France-Hawaii Telescope Legacy Survey; \citet{petrillo2017a}, the Kilo Degree Survey \citep{dejong2015a}; \citet{pourrahmani2018a}, the Hubble Space Telescope ACS $i$-band observations of the Cosmological Evolution Survey \citep[COSMOS;][]{capak2007a} field;
and \citet{jacobs2019a}, the Dark Energy Survey \citep[DES;][]{des2005a}.
All of these searches were performed on completed surveys.}

We decided to 
use only \textit{observed} data for lenses \textit{and} non-lenses in our training sample \ed{for partial deployment on DECaLS, which is near completion,} and have obtained encouraging results.  
We identify the known lenses 
in the Legacy Surveys and DES DR1.  
A catalog of known lenses in the Legacy Surveys is also necessary in order to identify new lens candidates. 
Both DECaLS and DES used DECam
(see D19).  
DES has $grizY$ observations with \rf{greater} depths \rf{(see Figure~\ref{fig:decals})} in the three bands common with Legacy Surveys.
Due to the paucity of lenses, we have used known strong lenses in all of Legacy Surveys, 
while in this paper we will focus on finding new lenses only in the DECaLS footprint.
The Master Lens Database\footnote{\url{http://admin.masterlens.org/index.php}} \ed{\citep{moustakas2012a}}, which contains hundreds of lensing events up to 2016,
\ed{provided the initial list for the lens training sample.}
We have since added several hundred more lenses \ed{and lens candidates} from more recent publications \citep{carrasco2017a, diehl2017a, pourrahmani2018a, sonnenfeld2018a, wong2018a, jacobs2017a, jacobs2019a}.
In total we have identified $\sim$~700 \ed{previously known} lenses or \ed{lens candidates} in the Legacy Surveys and DES. 
A number of these systems were discovered spectroscopically or 
through imaging with better seeing \rf{and/or greater depth} than the Legacy Surveys and DES.
Some of them therefore have sub-arcsecond deflection angles 
\rf{and/or lensed sources fainter than can be clearly seen from the DECaLS and DES observations}.   
Through human inspection, we deem \rf{613} as discernible lenses in the Legacy Surveys (199) and DES (414) footprints. 
For the lenses in the DES footprint, we only include $grz$ bands.
We also assemble 13,000 non-lens image cutouts from the Legacy Surveys, 
all with at least three passes in each of the $grz$ bands.
Of these, 5000 are galaxies categorized as DEV or COMP in D19 (see \S~\ref{sec:observations}), which are elliptical galaxies, and another 5000 of all types of galaxies. 
For both cases, we apply a $z$-band magnitude cut of 22.5~mag. 
Given that on average we expect one strong lens in $\mathcal{O} (10^4)$ galaxies \citep[e.g.,][]{oguri2010a},
incidental inclusion of a lens or two in these \rf{randomly selected} galaxies is not a significant concern.  

The reason for including non-elliptical galaxies is to provide more non-lens configurations for the neural net.  
Two of the co-authors have also selected another 3,000 non-lenses by eye so as to cover as many non-lens configurations as possible, 
 especially cases that can potentially be confused by the neural net.  
These include \rf{spiral galaxies of different sizes and spiral arm configurations, elliptical galaxies, galaxy groups, 
images having objects with different colors (typically a blue galaxy next to a red one), 
cosmic rays appearing in different bands (some of which have curved trajectories), unusual arrangements of galaxies or stars, 
and finally certain data reduction artefacts (Figure~\ref{fig:hand-select}).} 
Simulated non-lenses typically do not cover these scenarios.  

\vspace{5 mm}
\begin{minipage}{\linewidth}
\makebox[\linewidth]{
   \includegraphics[keepaspectratio=true,scale=0.48]{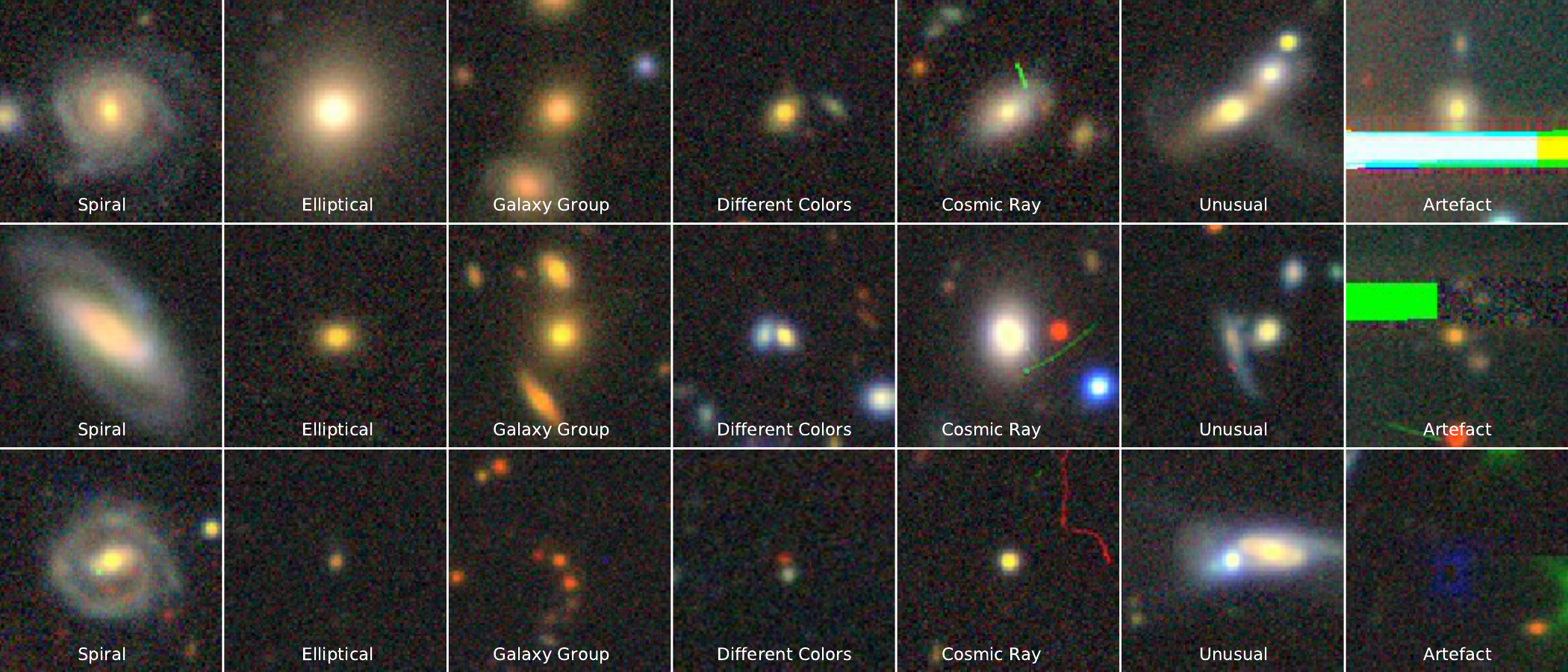}}
\captionof{figure}{ 
\rf{Examples of non-lens images in the training sample selected by eye.  
In order, the seven columns show the categories of (see text):
spiral galaxies,  
ellipitcal galaxies,  
galaxy groups,
galaxies with different colors, 
cosmic rays,  
unusual configurations, and finally  
data reduction artefacts. }}
\label{fig:hand-select}
\end{minipage}

\subsection{Residual Neural Networks}\label{sec:ResNet}

We have adopted the Residual Neural Network (ResNet) model of \citet[][L18]{lanusse2018a}\footnote{\url{https://github.com/McWilliamsCenter/CMUDeepLens}}, which used Theano\footnote{\url{http://deeplearning.net/software/theano/}} and Lasagne\footnote{\url{https://github.com/Lasagne/Lasagne}} libraries. 
We re-implemented their model in TensorFlow\footnote{\url{https://www.tensorflow.org/}}, 
in part because major development for Theano ceased after the 1.0 release on November 15, 2017.
We test the translated ResNet model using the simulated training set from the Strong Gravitational Lens Finding Challenge \citep{metcalf2018a} and have reproduced the results in L18, which was the winning entry for the Lens Challenge.
 The architecture of the model is described in detail in L18.

L18 has provided much guidance to our approach.  
At this stage we have left their architecture and hyperparameters unchanged, including the batch size (128), total number of training epochs (120), pre-processing of the images, 
and data augmentation (random rotation,  mirroring, and zooming within a range of [0.9, 1.0]; for details, see Section 3.3 of L18).
The lens and non-lens images in the training sample are cutouts with a dimension of 101~$\times$~101 pixels, following the specification in the Lens Challenge.  

We split the training sample into training, validation, and testing sets, with ratios of 70:20:10.
The sizes of our training, validation\rf{, and testing} sets are then, respectively, 9876 (423), 2818 (118), and \rf{1427 (72)} where the values in parentheses are the number of lenses. 
We set aside a testing set because we want to leave open the possibility of varying the architecture and hyperparameters to optimize the neural net's performance.  
We then train the ResNet 
on the supercomputer Cori\footnote{\url{https://www.nersc.gov/users/computational-systems/cori/}} at the National Energy Research Scientific Computing Center (NERSC)\footnote{\ed{\url{https://www.nersc.gov/}}}, using three Haswell computing nodes\footnote{\url{https://www.nersc.gov/users/computational-systems/cori/configuration/cori-phase-i/}}, one worker each. 
The 120 epochs of training took 17 hours.
The distributed training was accomplished by using Horovod\footnote{\url{https://github.com/horovod/}}.
Performing distributed training with deep \ed{(46 layers in this case; L18)} neural networks can be non-trivial.
We experimented with different numbers of decay epochs and found that with three workers, 
a decay epoch of 40 (i.e., the learning rate of the ResNet is decreased by a factor 10 every 40 epochs of training) works the best.  

\ed{The ResNet attempts to minimize the cross entropy loss function:}

\ed{\begin{equation}\label{eqn:loss}
    \displaystyle-\sum_{i=1}^{N} y_i \log \hat{y}_i+(1-y_i) \log (1-\hat{y}_i)
\end{equation}}

\noindent
\ed{where $y_i$ is label for the $i$th image (1 for lens and 0 for non-lens), and $\hat{y}_i \in [0,1]$ is the model predicted probability.}

\ed{While the loss function is monitored during the training process to determine the point of termination, the overall performance of the trained model is typically assessed by the Receiver Operating Characteristic (ROC) curve.  The ROC curve shows the True Positive Rate (TPR) vs. the False Positive Rate (FPR) for the validation set, where P(ositive) indicates a lens and N(egative), a non-lens.
With the definitions TP = correctly identified as a lens, False Positive = incorrectly identified as a lens, True Negative = correctly rejected, and False Negative = incorrectly rejected,}
\ed{\begin{equation*}
    \rm{TPR} = \frac{\rm{TP}}{\rm{P}} = \frac{\rm{TP}}{\rm{TP} + \rm{FN}}
\end{equation*}
}
\ed{and}
\ed{\begin{equation*}
    \rm{FPR} = \frac{\rm{FP}}{\rm{N}} = \frac{\rm{FP}}{\rm{FP} + \rm{TN}}
\end{equation*}
}

\noindent
\ed{The curve is generated by gradually increasing the threshold probability for a positive identification from 0 to 1.  
Random classifications will result in a diagonal line in this space with an area under the ROC curve (or AUC) equal 0.5. 
For a perfect classifier, AUC = 1.}

The decision of using three nodes was based on our experience with a smaller training set.  
We \ed{can significantly shorten} the training time by employing six or more nodes.  
Since the training set has a total of 9876~images,
with a batch size of 128 images and 3 workers, 
it takes 26 steps to complete one full training epoch.

In Figure~\ref{fig:loss-roc}, left panel, 
we show how the cross entropy loss functions vary as training progresses.  
For the validation set, we show the value at every epoch.  
For the training set, the cross entropy was reported for every step, which we have boxcar smoothed with a window size of 26.
\ed{As L18 also noted, the loss function (Equation~\ref{eqn:loss})} and the AUC for the validation set both plateau well within 120 epochs of training.
Since the model has performed well, we have left the architecture and hyperparameters in L18 unchanged and moved directly to deployment.  
Thus so far we have not used the validation set, or the testing set, for training.

\noindent%
 \begin{minipage}{\linewidth}
 \makebox[\linewidth]{
   \includegraphics[keepaspectratio=true,scale=0.7]{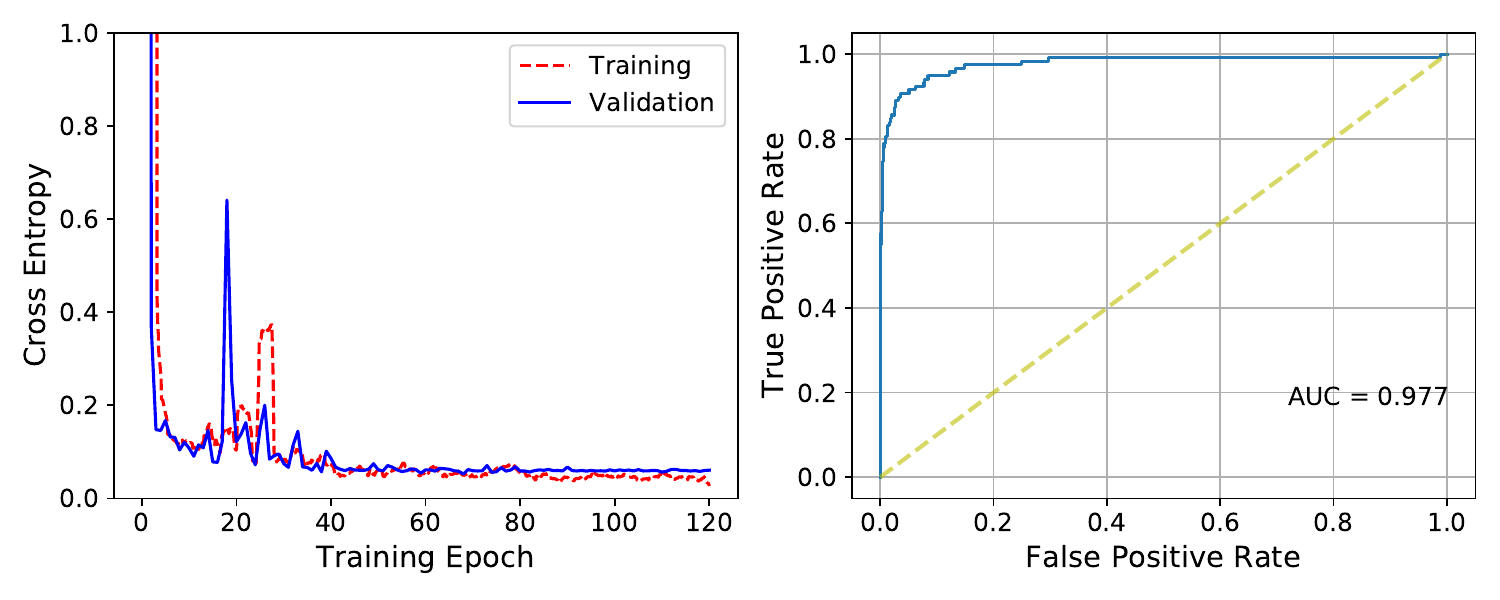}}
 \captionof{figure}{ 
\textbf{Left}: The cross entropy loss functions for the training and validation sets over 120 epochs.
\textbf{Right}: The receiver operative characteristic (ROC) curve for the validation set with the area under the curve (AUC) = 0.98.}\label{fig:loss-roc}
 \end{minipage}



We achieve an AUC of 0.98 for the validation set (Figure~\ref{fig:loss-roc}, right panel).  
Even though our training and validation sets contain far fewer lenses, our AUC matches the performance on simulated data in L18. 


%% file: results.tex
\subsection{Inference and Lens Candidates}\label{sec:candidates}

We apply our trained ResNet model to 5.7 million DEV and COMP type galaxies in DECaLS with at least three passes in each of the three bands ($grz$) and $z$-band magnitude $\leq$ 20.0. 
This magnitude cut was chosen because it 
includes \ed{92\%} of the known lenses in the Legacy Surveys and results in a manageable number of images for human inspection.   

\rf{Four co-authors (AP, CS, MD, and VR) have inspected $\sim 50,000$ cutout images that receive a probability $\geq 0.9$, 
evenly split.  They select candidates according to these criteria, erroring on the generous side:} 
small blue galaxy/galaxies (red galaxies are rare but certainly acceptable) next to the red galaxy/galaxies at the center that 
    \begin{itemize}
        \item are typically 1 - 5$\twopr$ away   
        \item have low surface brightness
        \item curve toward the red galaxy/galaxies
        \item have counter/multiple images with similar colors (especially in Einstein-cross like configuration)
        \item are elongated (including semi- or nearly full circles)

    \end{itemize}

\noindent
Typically, most candidates do not have all these characteristics. 
In general, the greater the number of characteristics listed above an image has, the higher they are ranked by humans.
\rf{A fifth co-author (XH), in consultation with another co-author (RG), examines all these ``first pass" candidates and assigns the grades of A, B, and C, and discards the rest:}

\begin{itemize}
    \item Grade~A: We have a high level of confidence of these candidates.  
    Many of them have one or more prominent arcs, usually blue.
    The rest have one or more clear arclets, sometimes arranged in counter-image configurations with similar colors (again, typically blue).  \ed{However, there are clear cases with red arcs.}
    
    \item Grade~B:   
    They have similar characteristics as the Grade A's.  
    For the cutout images where there appear to be giant arcs they tend to be fainter than those for the Grade A's.  
    Likewise, the putative arclets tend to be smaller and/or fainter, or isolated (without counter images).
    
    \item Grade~C: They generally have features that are even fainter and/or smaller than what is typical for the Grade~B candidates, but that are nevertheless suggestive of lensed arclets.  
    They are usually without counter images, 
    \ed{except for a few cases.} 
    In almost all cases, if these are indeed lensing systems, the deflection angles are comparable to or only slightly larger than the seeing.
\end{itemize}

\noindent
\ed{For Grade~B and C candidates, we have included a few cases where it is difficult to judge whether it is a lensing event vs. a coincidental placement of galaxies, a spiral galaxy, or a ring galaxy.} 
In total we have identified \lenstot candidates :\lensA A's, \lensBnew B's, and \lensCnew C's, 
listed in Tables~\ref{tab:grade-A}, \ref{tab:grade-B}, and \ref{tab:grade-C} and shown in Figures~\ref{fig:grade-a}, \ref{fig:grade-b}, and \ref{fig:grade-c}.
\rf{The locations of the candidates in the sky are shown in Figure~\ref{fig:candidates}.}

 \begin{minipage}{\linewidth}
 \makebox[\linewidth]{
   \includegraphics[keepaspectratio=true,scale=0.47]{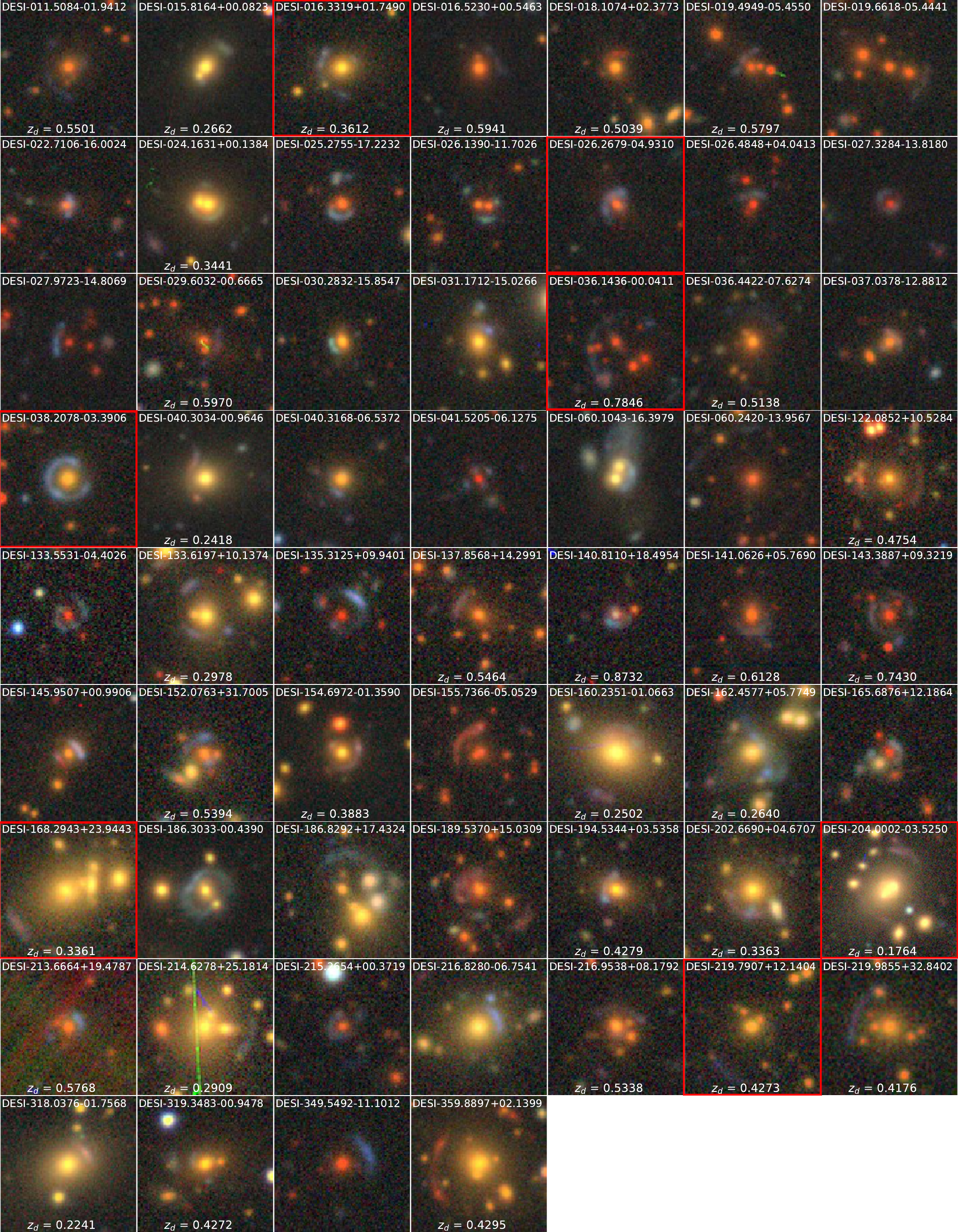}}
 \captionof{figure}{ 
 The \ed{\lensA} Grade A lens candidates \ed{arranged in ascending RA.}  
 Each image is 101 pixels $\approx 26.2 \twopr$ on the side, \ed{with N up and E to the left.}
 The exception \ed{is} DESI-204.0002-03.5250,
 which is 151 pixel $\approx 40.3\twopr$ on the side.
\rf{The seven images with a red rim are later found to be known lenses 
(see text)}.  
\rf{These seven lenses were not included in our training sample, 
and they happen to be in the selected subset of the Legacy Surveys data of 5.7 million
elliptical galaxies.}
 }\label{fig:grade-a}
 \end{minipage}

\noindent%
 \begin{minipage}{\linewidth}
 \makebox[\linewidth]{
   \includegraphics[keepaspectratio=true,scale=0.35]{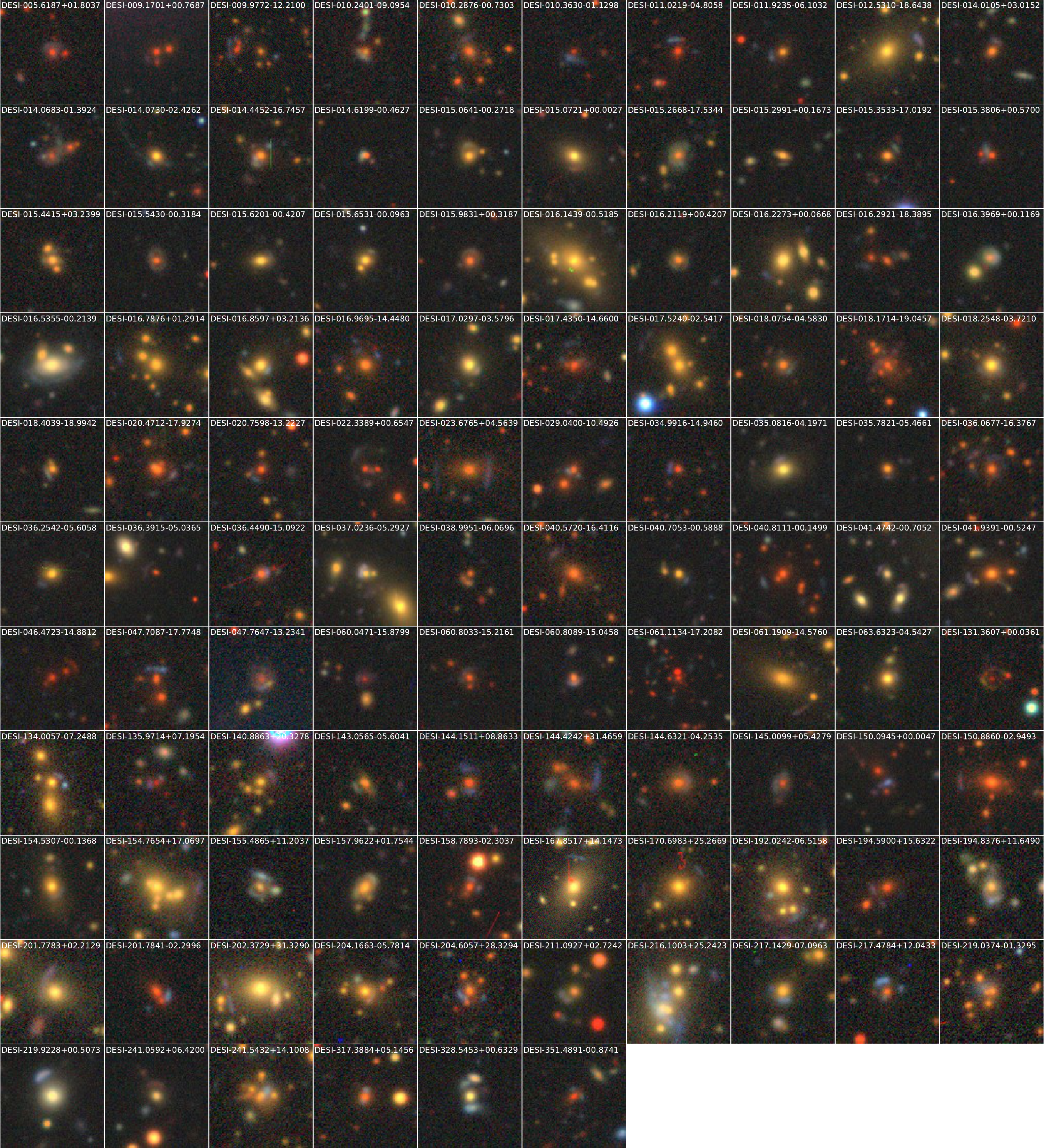}}
 \captionof{figure}{ 
 The \ed{\lensBnew} Grade B Lens Candidates.  
 Each image is 101 pixels $\approx 26.2 \twopr$ on the side, with N up and E to the left.  
 The exceptions are 
DESI-009.9772-12.2100, 
 DESI-061.1134-17.2082, and
 \ed{DESI-167.8517+14.1473}, which are 151 pixels $\approx 40.3\twopr$ on the side.
 }\label{fig:grade-b}
 \end{minipage}

\noindent%
 \begin{minipage}{\linewidth}
 \makebox[\linewidth]{
   \includegraphics[keepaspectratio=true,scale=0.31]{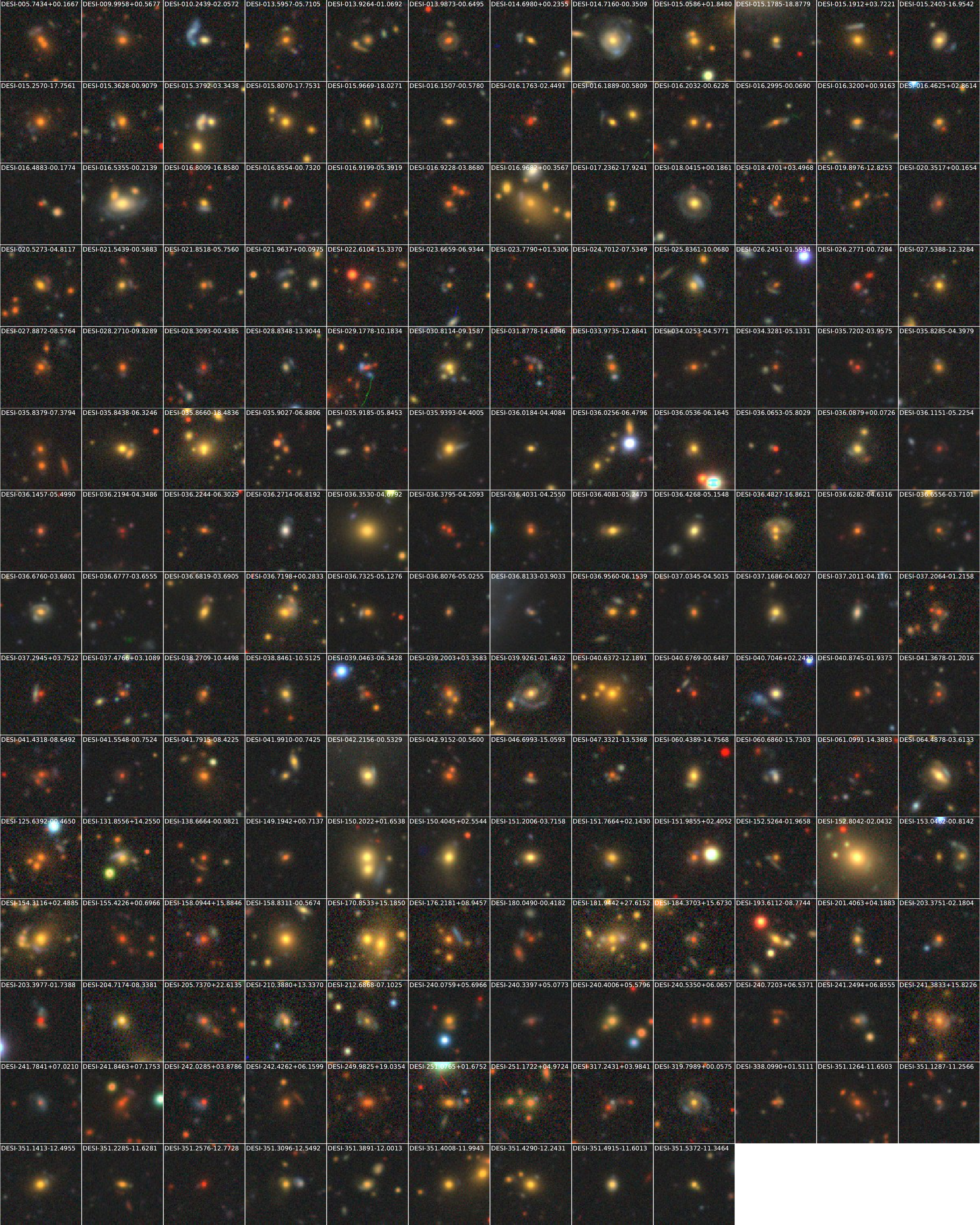}}
 \captionof{figure}{ 
 The \ed{\lensC} Grade C Lens Candidates.  
 Each image is 101 pixels $\approx 26.2\twopr$ on the side, with 
N up and E to the left.}\label{fig:grade-c}
 \end{minipage}

\begin{minipage}{\linewidth}
\makebox[\linewidth]{
  \includegraphics[keepaspectratio=true,scale=0.48]{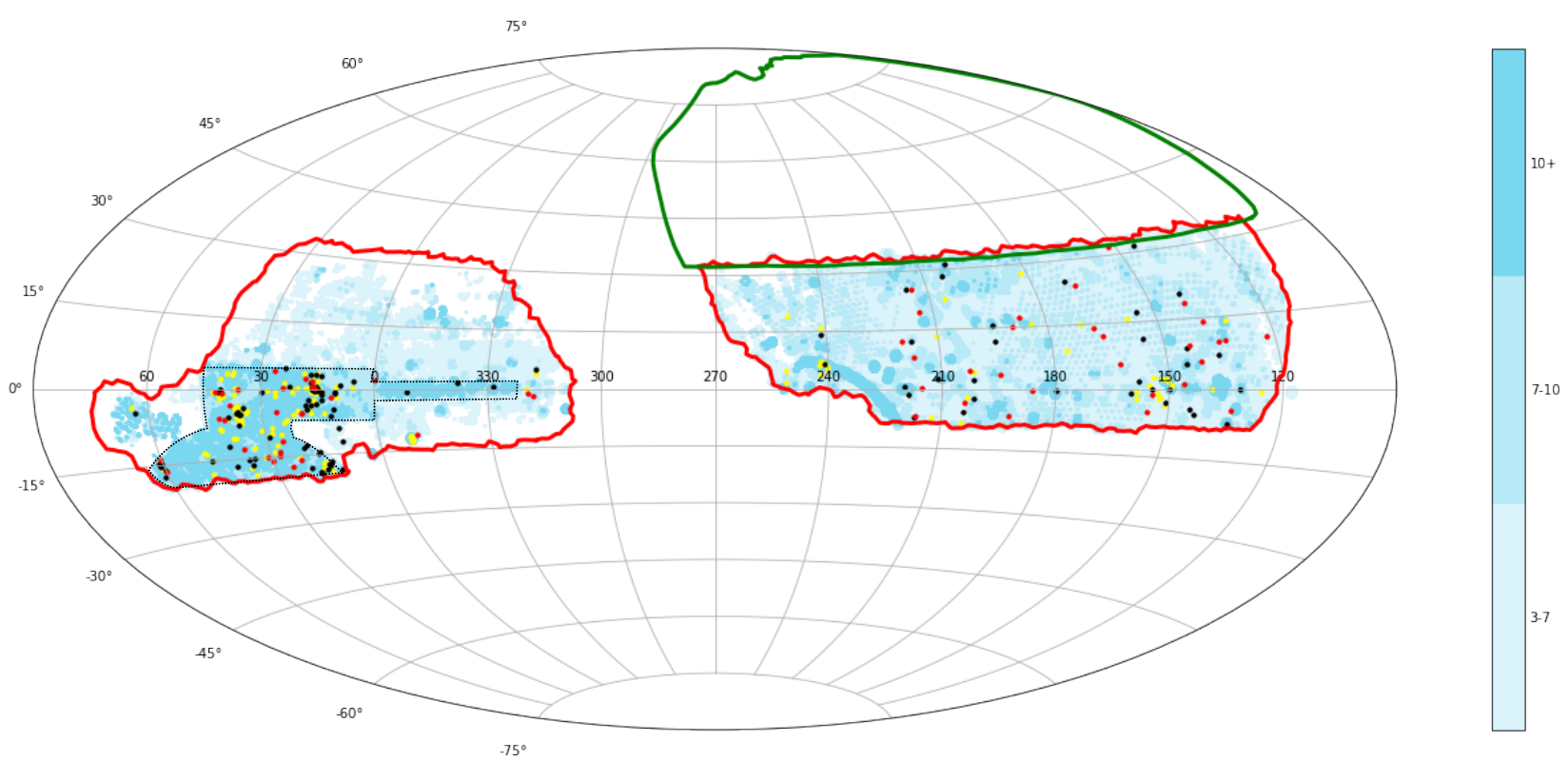}}
   \captionof{figure}{\rf{
The new candidate lensing systems identified in the DECaLS region are shown as red (Grade A), black (Grade~B), and yellow (Grade~C) circles.
The footprint of DECaLS is highlighted by red outlines and the $z$-band depth is represented by different shades of blue (for more details, see the caption for Figure~\ref{fig:decals}).
  }}
  \label{fig:candidates}
\end{minipage}



\input{table-grade-A.tex}

\input{table-grade-B.tex}

\input{table-grade-C.tex}

We have checked our candidates against the \HST Source Catalog (HSCv3)\footnote{\url{https://mast.stsci.edu/portal/Mashup/Clients/Mast/Portal.html}},
and found six \ed{known lenses} among the Grade A candidates: 
\ed{DESI-016.3319+01.7490 and DESI-026.2679$-$04.9310 \citep{stark2013a}, 
DESI-036.1436$-$00.0411 \citep{gladders2003a}, 
DESI-168.2943+23.9443 \citep{kubo2009a},
DESI-204.0002$-$03.5250 and DESI-219.7907+12.1404
\citep[SDSS DR12 BCG;][]{sharon2019a}.}
\rf{A seventh, DESI-038.2078-03.3906 \citep{stark2013a}, 
was pointed out to us during the peer review process 
(by Chien-Hsiu Lee at NOAO after the paper was posted on the arXiv).}  
These are not in our training sample, and shown with a red rim in Figure~\ref{fig:grade-a}.  
This leaves the number of new Grade~A candidates as \lensAnew, and the total number of new lens candidates, \lenstotnew.

We have found at least \ed{13 new cluster/group}  
scale strong lenses: 
\ed{DESI-019.6618-05.4441, 
DESI-060.2420-13.9567,
DESI-167.8517+14.1473
DESI-219.9855+32.8402, 
and DESI-359.8897+02.1399 (with a giant red arc) among Grade A, and 
DESI-009.9772-12.2100,
DESI-018.1714-19.0457, 
DESI-022.3389+00.6547, 
DESI-023.6765+04.5639, 
DESI-061.1134-17.2082, 
DESI-154.7654+17.0697, 
DESI-202.3729+31.3290,
and DESI-216.1003+25.2423
}
among Grade B candidates.

Among the hundreds of galaxy scale candidates,
there are many notable lensing events.  
We especially would like to highlight: \textbf{DESI-135.3125+09.940}, a system having a lens with $g - r = 3.3$, likely indicating a high redshift \citep[e.g.,][]{jacobs2019a}; \textbf{DESI-041.5205-06.1275}, a nearly perfect Einstein Cross; 
\rf{and \textbf{DESI-155.7366-05.0529}, with a red arc, which
would have a redshift of $\sim 4$ if it is a Lyman-break galaxy.}

We would like to emphasize the importance of the Grade B and C candidates.
The ``rediscovered lenses" outside the training sample are in the A category (seven out of \lensA). 
This is not a surprise: typically having brighter arcs with larger deflection angles, these systems are comparatively easy to find.
Higher redshift lensing systems from ground-based surveys are likely not in the Grade~A category but in B or C.
The current known lensing sample mostly consists of luminous elliptical galaxies at redshifts from approximately 0.4 to 0.8 \citep[e.g.,][]{brownstein2012a, wong2018a}.
Our lens candidates are fainter, and mostly have optical and infrared colors consistent
with $z > 0.8$ \citep[e.g.,][]{jacobs2019a}.  
Higher lens redshifts significantly increase the power relative to lower redshift samples for constraining the mass function of low-mass CDM halos, due to the greater optical depth for perturbations by low-mass halos associated with a longer path length along the line of sight \citep{despali2018a, Ritondale2019a}.  
In addition, the lensed sources will tend to have higher redshifts than in known lensing systems as well.

Finally, we checked our candidate list against the spectroscopic database from SDSS~I and II \citep{york2000a}, SDSS~III/BOSS \citep{eisenstein2011a}, and SDSS~IV/eBOSS \citep{blanton2017a} and found \lensz matches, which is slightly greater than one third of all candidates.
The available redshifts are included in Tables~\ref{tab:grade-A}, \ref{tab:grade-B}, and \ref{tab:grade-C}.

The completeness is difficult to estimate at this point, 
even just for elliptical galaxies  
because 1) the data reduction for the Legacy Survey has not completed (recall for this deployment, we have only included images with at least 3 passes in each band)
and 2) we have not run inference on the REX category. 

\subsection{Probability Bins Lower than 0.9}\label{sec:low-p}

It is notable that there are typically many more candidates with probability greater than 0.9 than with probabilities between 0.8 and 0.9.  
In a small testing inference run that covers $\approx$~4\% of the DECaLS footprint, 
we have examined and found lens candidates with  probability \ed{$<0.9$}.  
The yield typically rapidly diminishes with lower probabilities.
As stated earlier, 
for the deployment on galaxies typed DEV and COMP, 
we impose a magnitude cut at $z \leq 20.0$~mag.  
For our small test inference run, we included all objects with $z \leq 22.5$~mag.  
From that run, we have found one Grade~B lens (DESI-135.9714+07.1954) with a $z$-band magnitude of 20.87.

Given that a strong majority of the best lens candidates are from the probability $> 0.9$ bin for the categories of DEV and COMP,
in this paper we focus on this subset for human inspection.

A rough estimate of completeness can be performed by checking how many lensing systems from the 
\rf{validation (118) and testing (72) set would be ``re-discovered."  
This depends on the threshold.
Seventy seven of the 190 lenses in the validation and testing sets received a probability of $> 0.9$, or about 40\%. 
The precision (TP/(TP+FP)) vs. recall (TP/(TP + FN), or completeness) curve for the validation and testing sets, with probabilty threshold values marked, are shown in Figure~\ref{fig:precision-recall}.}

\begin{minipage}{\linewidth}
\makebox[\linewidth]{
  \includegraphics[keepaspectratio=true,scale=0.1]{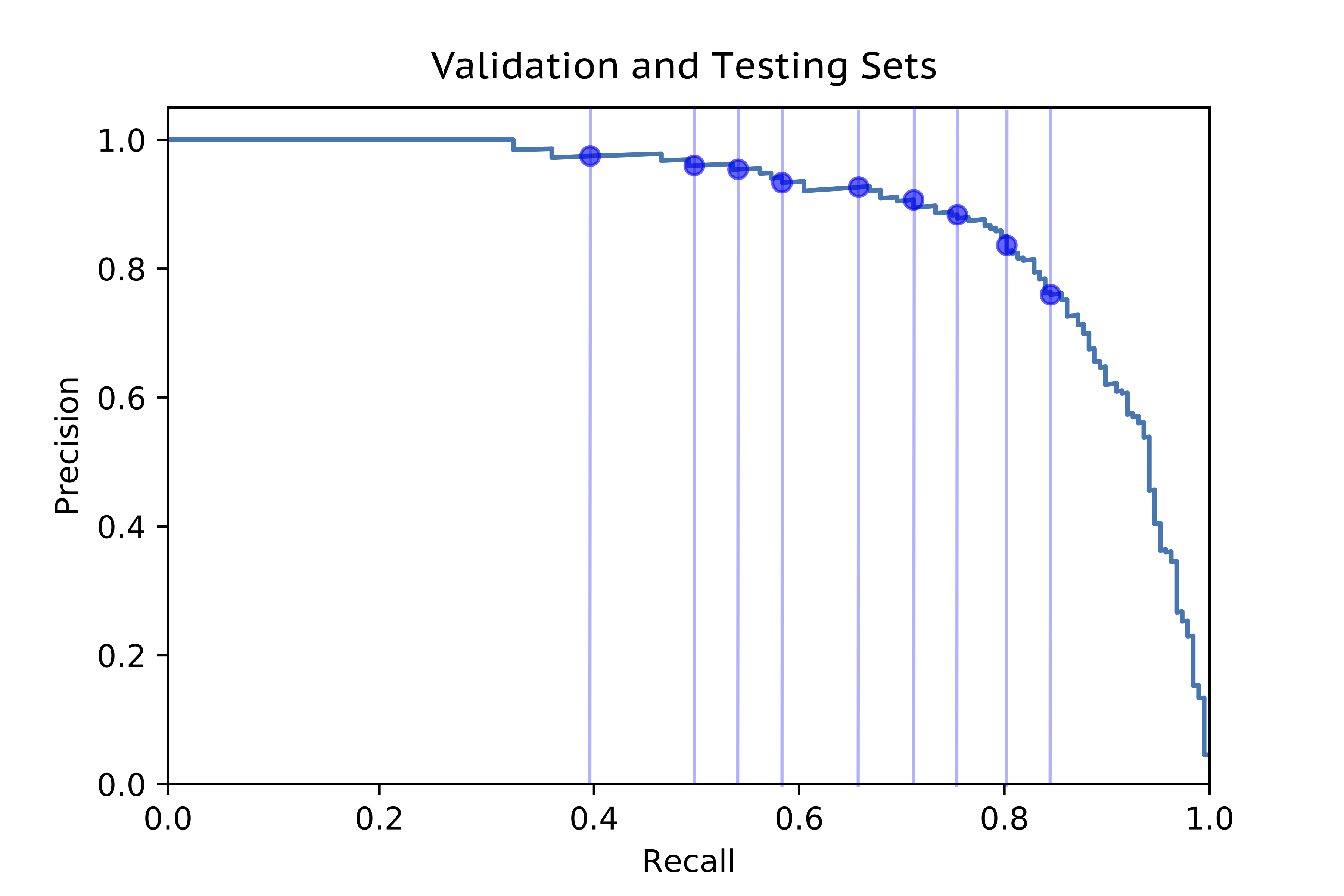}}
  \captionof{figure}{\rf{
 The precision-recall curve for the validation and testing sets.  
 The blue points from left to right correspond to probability threshold values from 0.9 to 0.1 with an interval of 0.1.
  }}
  \label{fig:precision-recall}
\end{minipage}

The implication seems to be that there are hundreds more lenses to be discovered in lower probability bins. 
We can already confirm there are good lens candidates with probability below 0.9 based on visual inspection for a small subset of the data.
However, we caution against a simple forecast based on these percentages.  
The images in the training sample that receive low probability often are less obvious lenses for the human inspector.
In fact for the next round of training, we would remove some of them from the training sample.

\rf{  
To demonstrate, in Figure~\ref{fig:train-lenses-remove}, 
we show four examples of DECaLS images of lensing systems in the current training sample that we would consider for removal.
Without prior knowledge, these images could not have been determined to be lenses by visual inspection.  
The first three are from the Strong Lensing Legacy Survey \citep[SL2S;][]{cabanac2007a, more2012a}.  
This survey was carried out on the Canada-France-Hawaii Telescope (CFHT) in $urgiz$ bands and the lensing systems were found from both their WIDE and DEEP components. 
They ran semi-automated searches \citep{more2012a} on their $g$-band images, 
which are $\sim 1.5$~mag deeper than the $5\,\sigma$ $g$-band depth of DECaLS, 
and has a mean seeing of $0\twopr.78$.
The DEEP fields were almost 10 times deeper.  
Therefore some of the blue arcs in the SL2S images are barely visible in their DECaLS counterparts, 
either due to the shallower $g$-band depth (Figure~\ref{fig:train-lenses-remove}, Panels (a) and (b)) and/or the comparatively inferior seeing of DECaLS (Figure~\ref{fig:train-lenses-remove}, Panel (c)).
}
\rf{
The fourth example in Figure 3 (Panel (d)) comes from the Hyper Suprime-Cam Subaru Strategic Program \citep[HSC-SSP;][]{aihara2018a}.  
Observed in $grizy$ bands, this survey also has greater depth and better seeing than DECaLS.  
As a result, some of the blue arcs seen in the HSC images \citep{sonnenfeld2018a} are barely visible in DECaLS.
}

\rf{
From the experience of this deployment, we believe including images like these in the training sample has resulted in a number of images with high probability that cannot be unambiguously classifiied by visual inspection.  
These images significantly decrease the efficiency of human inspection without clear improvement for completeness.  
This intuition seems to be correct from our preliminary results using an updated training sample.  
In the future, we will also experiment with getting cutout images directly from the SL2S and HSC surveys.
}

\vspace{5mm}
 \begin{minipage}{\linewidth}
 \makebox[\linewidth]{
   \includegraphics[keepaspectratio=true,scale=0.6]{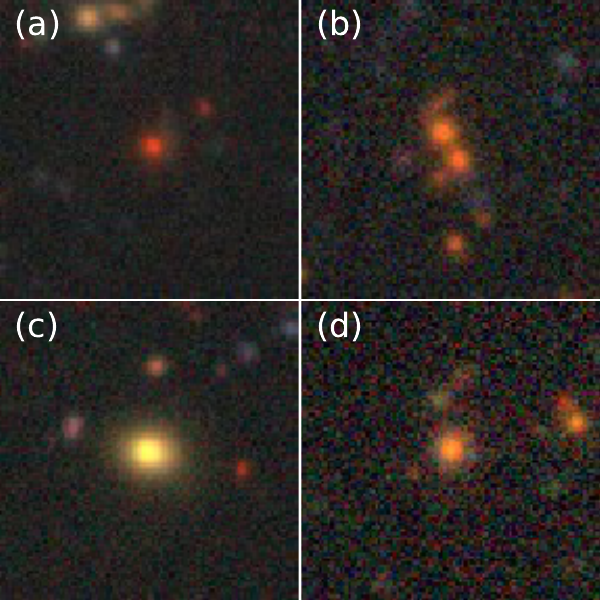}}
 \captionof{figure}{ 
\rf{Four examples of lensing systems where the blue arcs are very hard to tell, if at all, from DECaLS images,
either due to their shallower $g$-band depth and/or slightly worse seeing (the deflection angle of the system in Panel (d) is $\sim0\twopr.5$). 
The systems shown in Panels (a) - (c) are from the Strong Lensing Legacy Survey, 
and the one in (d), the Hyper Suprime-Cam Subaru Strategic Program (see text).}
 }\label{fig:train-lenses-remove}
 \end{minipage}



\subsection{\rf{Human Inspection Efficiency}}\label{sec:purity}

Below we briefly discuss the \rf{human inspection efficiency} of the
ResNet results thus far.
In total we have examined $\sim 50,000$ objects.  
On average one in 150 of the objects \rf{with probability from the ResNet model} is \rf{deemed} a lens candidate \rf{through human inspection}.  
The Legacy Surveys data is catalogued by Tractor and organized in folders, with each folder corresponding to one degree of RA on the sky.
The efficiency of our trained ResNet is highly uneven.
The number of ResNet-recommended objects 
per folder in the probability $>0.9$ bin vary from under 200 to over 3000.
We have examined folders with both small and large numbers of objects.
In general the folders with lower numbers of objects have higher purity.

\rf{  
We illustrate this trend in Figure~\ref{fig:objs-recs}, by using the number of passes for $z$-band as a proxy for depth (as in Figure~\ref{fig:decals}).
The left panel (orange columns) shows that while $\sim 5\%$ of the objects in the 10+ pass bin receive probability $>0.9$ (``recommendations"), only $\sim 0.1\%$ in the 3 - 7 passes bin do.
There are certainly more lenses to be found in the deeper regions, but not 50~times more likely. 
Human inspection in fact reveals that a much lower percentage of the recommendations from the deepest bin are lens candidates (0.35\%), compared with the shallower ones (0.45\% for 7 - 10 passes and 1.61\% for 3 - 7 passes; Figure~\ref{fig:objs-recs}, left panel, gray columns). 
The ResNet's preference for images with greater depth is also shown in the middle panel of Figure~\ref{fig:objs-recs}:
while over 50\% of the objects (blue column) are from the shallowest bin (3 - 7 passes), the fraction of recommendations is just below 20\% (orange column); 
in contrast, objects from the deepest bin (10+ passes) only account for $\sim 5\%$ of the objects and yet over 80\% of the recommendations are from this bin (again, blue and orange columns, respectively).  
}

\rf{
This is not a surprise.  
Since lensed sources typcially are faint, most of the lenses in our training sample are in regions of greater depth (Figure~\ref{fig:lenses-nonlenses-training}, left panel and Figure~\ref{fig:objs-recs}, right panel, violet columns). 
The non-lenses in the training sample are mostly selected from shallower regions (Figure~\ref{fig:lenses-nonlenses-training}, right panel and Figure~\ref{fig:objs-recs}, right panel, yellow columns).
Such a training sample leads the ResNet to assign probabilities not only \rff{based} on how likely an image contains a lensing system  but also its depth. 
Note that in Figure~\ref{fig:objs-recs}, Panel~(c), 
there is a small fraction of lenses in the training sample with fewer than three passes.  
Due to the small number of known lenses, we wanted to include as many lenses as possible.
}

\rf{
We have found that the way to address this confusion is to include deep non-lens images.
We will perform a full search for DR8, which was completed during the peer review process for this paper, 
using a larger and more statistically representative training sample
(for more details,
see \S~\ref{sec:better-model})}.

\vspace{0.2 in}

\begin{minipage}{\linewidth}
\makebox[\linewidth]{
\includegraphics[width=.306\linewidth]{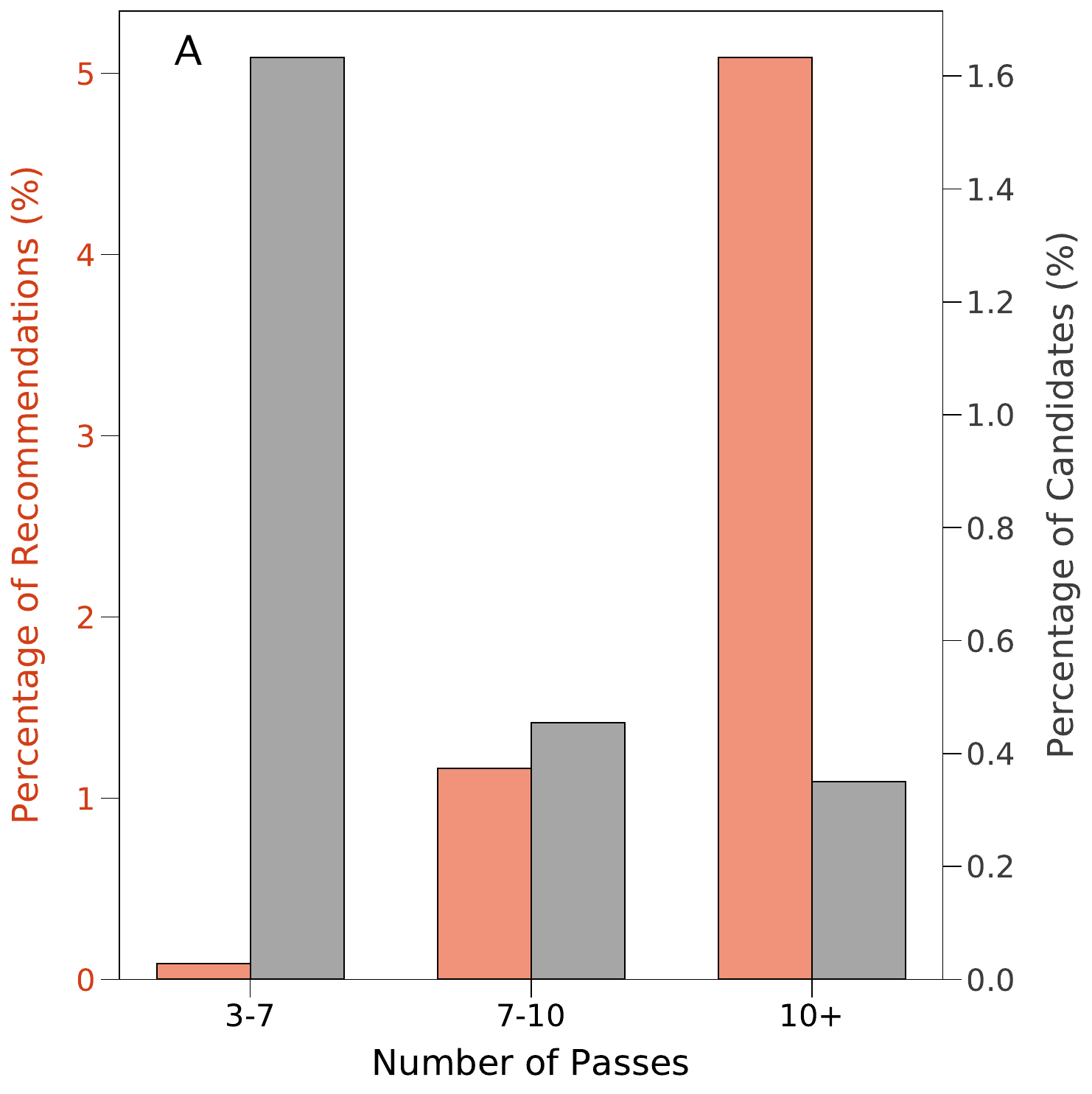}\hfill
\includegraphics[width=.306\linewidth]{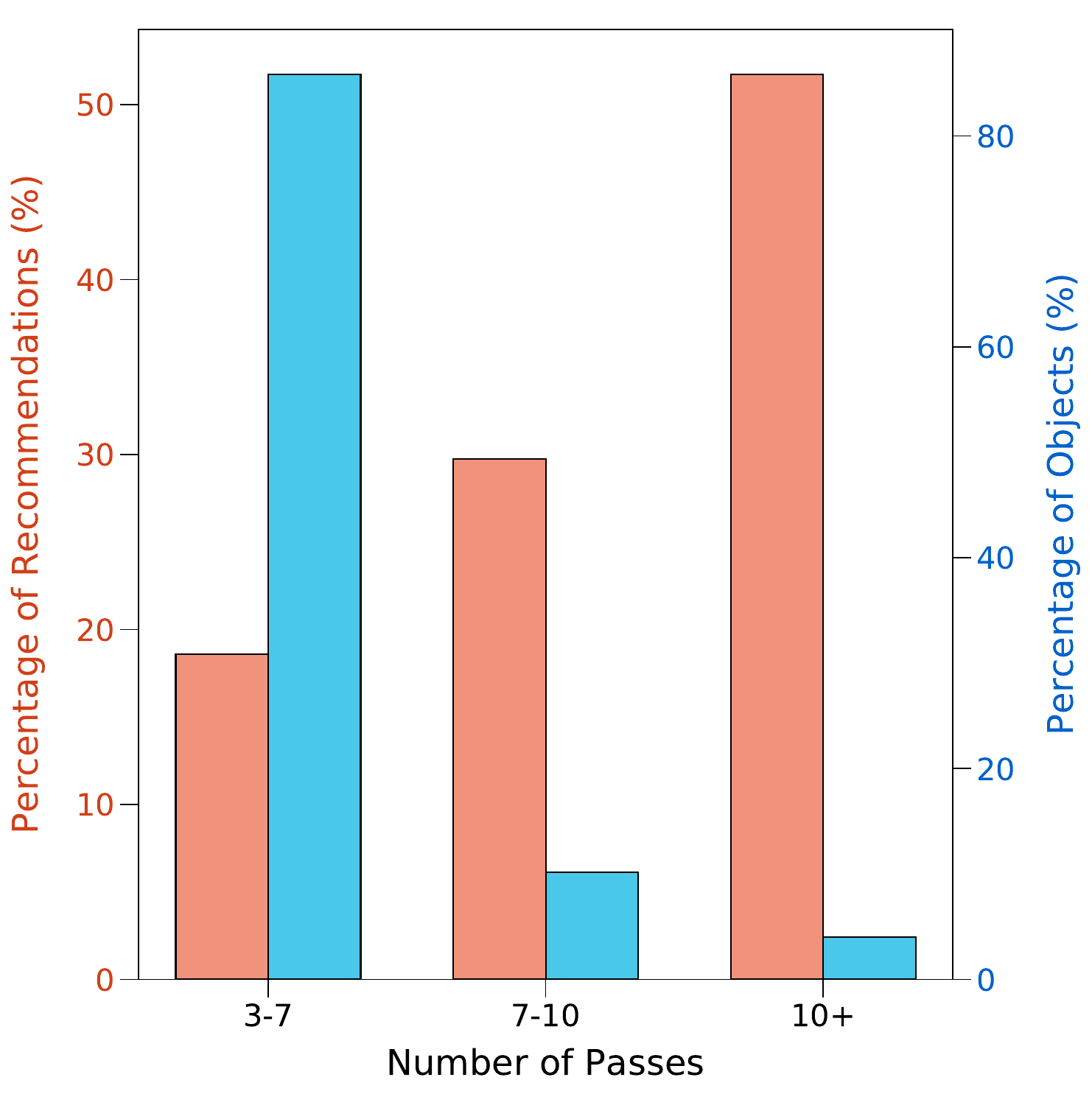}\hfill
\includegraphics[width=.385\linewidth]{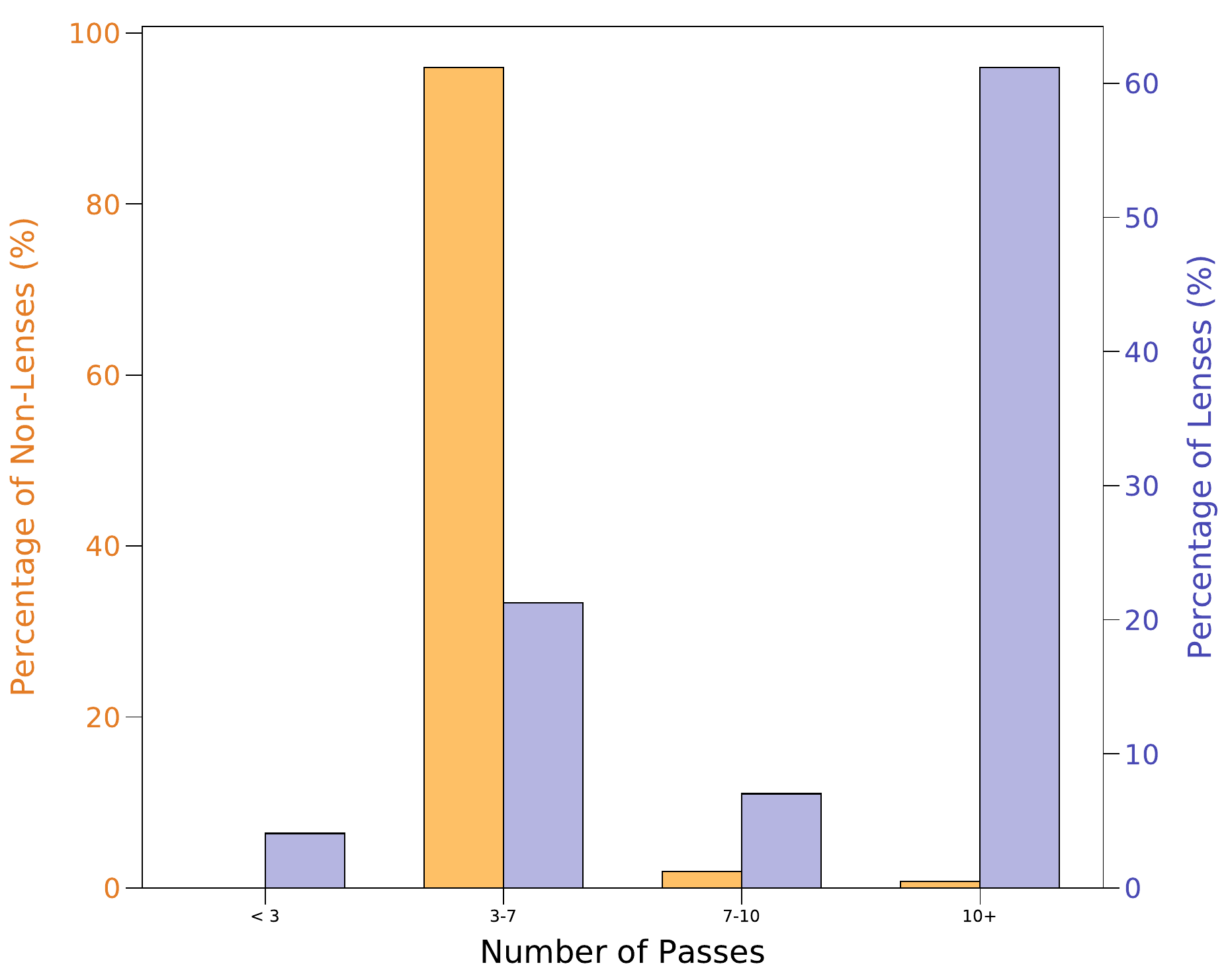}}	
	\captionof{figure}{
\rf{	\textbf{Left:} The orange columns (left $y$-axis) show the percentages of objects given a greater than 0.9 probability by our ResNet model (or ``recommendations") for the three bins of $z$-band depth.  
	The gray columns (right $y$-axis) show the percentages of ResNet recommendations that are selected as lens candidates through human inspection.
	\textbf{Center:} The orange columns (left $y$-axis) show the \rff{percentages} of ResNet recommendations for each bin of $z$-band depth.
	The blue columns (right $y$-axis) show the percentage of objects in each bin. 
These two panels show that the ResNet favors images with greater depth whether they contain a lensing system or not.
This is a consequnce of the composition of our current training sample,  
as shown in the last panel.
\textbf{Right:}  The yellow (left $y$-axis) and violet (right $y$-axis) columns show the \rff{percentages} of lenses and non-lenses in the training sample, respectively.  
Overall the images containing lensing systems tend to be much deeper, 
though a small fraction of them are shallower (see text).
It is not a surprise that with such a training sample, we get the distributions of the recommendations and candidates relative to the depth seen in the first two panels. 
	}}
  \label{fig:objs-recs}
\end{minipage}

\begin{minipage}{\linewidth}
	\begin{minipage}[t]{0.45\textwidth}
  	\includegraphics[keepaspectratio=true,scale=0.2]{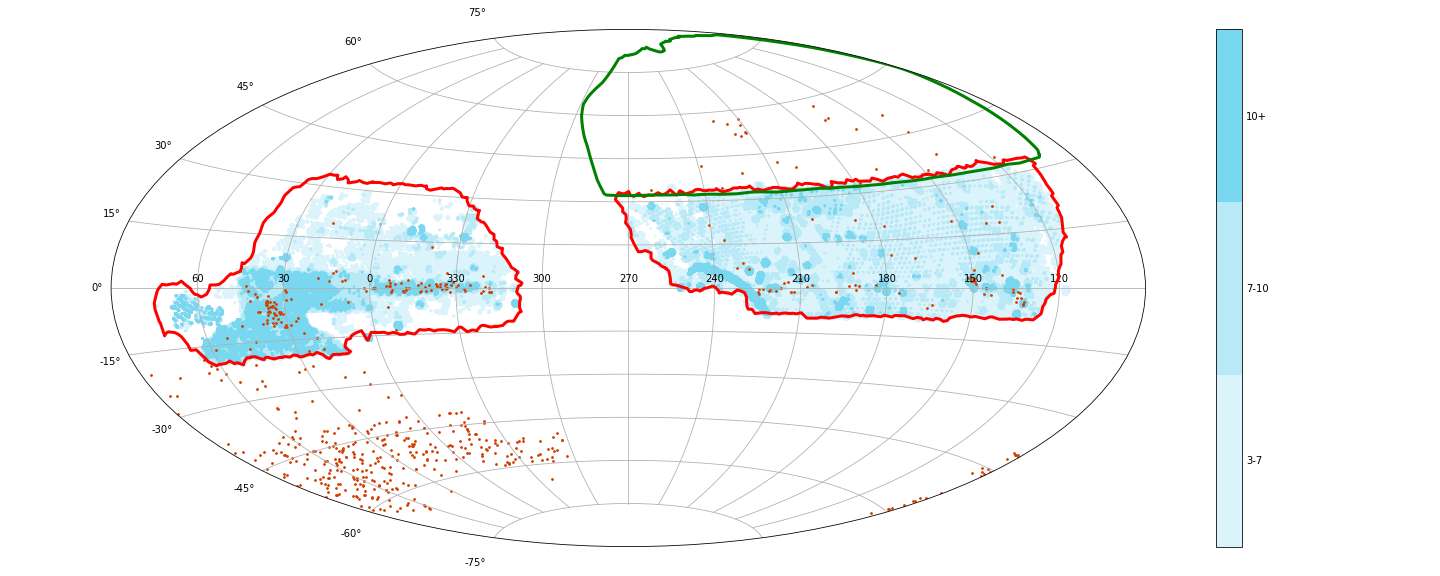}  
	\end{minipage}
	\begin{minipage}[t]{0.45\textwidth}
    \includegraphics[keepaspectratio=true,scale=0.2]{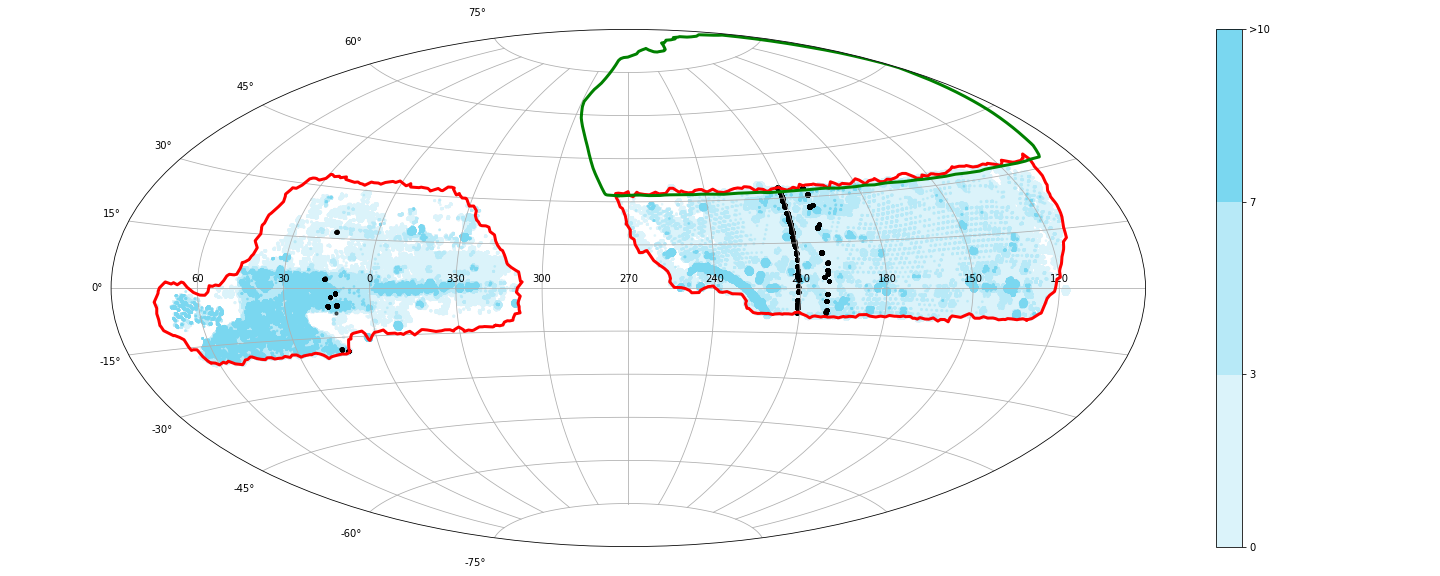}
	\end{minipage}
	\captionof{figure}{\rf{Lenses (red dots, left panel) and non-lenses (black dots, right panel) in our training sample, 
against the background of the depth map of DECaLS DR7 
(see the caption for Figure~\ref{fig:decals}).
The lenses outside the DECaLS footprint are from DES, all with similar depth (10+ passes in $z$ band).}}
\label{fig:lenses-nonlenses-training}
\end{minipage}

%% file: table-grade-A.tex
\clearpage 
   \startlongtable 
   \begin{deluxetable*}{lccccccr} 
  \tabletypesize{\footnotesize} 
   \tablecaption{ 
   Grade A Candidates\label{tab:grade-A}} 
   \tablehead{ 
   \colhead{Name} & 
   \colhead{Type} & 
    \colhead{mag\_g} &  
   \colhead{mag\_r} &  
   \colhead{mag\_z} &  
   \colhead{Probability} & 
   \colhead{$z$} & 
   \colhead{Survey}} 
   \startdata 
DESI-011.5084-01.9412 & DEV & 20.94 & 19.19 & 17.97 & 0.936 & 0.5501 & BOSS\\ 
DESI-015.8164+00.0823 & DEV & 19.64 & 18.08 & 17.21 & 1.000 & 0.2662 & SDSS\\ 
DESI-016.3319+01.7490 & DEV & 19.95 & 18.17 & 17.27 & 0.987 & 0.3612 & BOSS\\ 
DESI-016.5230+00.5463 & DEV & 21.52 & 19.66 & 18.25 & 1.000 & 0.5941 & BOSS\\ 
DESI-018.1074+02.3773 & DEV & 21.27 & 19.35 & 18.12 & 0.948 & 0.5039 & BOSS\\ 
DESI-019.4949-05.4550 & DEV & 21.41 & 19.66 & 18.37 & 0.984 & 0.5797 & BOSS\\ 
DESI-019.6618-05.4441 & DEV & 22.30 & 20.38 & 18.99 & 0.969\\ 
DESI-022.7106-16.0024 & COMP & 19.44 & 18.64 & 17.61 & 0.948\\ 
DESI-024.1631+00.1384 & COMP & 18.83 & 17.18 & 16.29 & 0.998 & 0.3441 & SDSS\\ 
DESI-025.2755-17.2232 & DEV & 20.43 & 18.96 & 17.62 & 0.989\\ 
DESI-026.1390-11.7026 & COMP & 20.09 & 18.39 & 17.00 & 0.961\\ 
DESI-026.2679-04.9310 & COMP & 20.75 & 19.67 & 18.43 & 0.998\\ 
DESI-026.4848+04.0413 & DEV & 22.29 & 20.78 & 19.05 & 0.927\\ 
DESI-027.3284-13.8180 & DEV & 22.13 & 20.95 & 19.22 & 0.998\\ 
DESI-027.9723-14.8069 & DEV & 21.62 & 20.51 & 18.90 & 1.000\\ 
DESI-029.6032-00.6665 & DEV & 21.05 & 19.45 & 17.94 & 0.918 & 0.5970 & SDSS\\ 
DESI-030.2832-15.8547 & DEV & 21.12 & 19.26 & 18.16 & 0.997\\ 
DESI-031.1712-15.0266 & DEV & 20.08 & 18.35 & 17.40 & 0.996\\ 
DESI-036.1436-00.0411 & DEV & 21.67 & 20.33 & 18.78 & 0.962 & 0.7846 & eBOSS\\ 
DESI-036.4422-07.6274 & DEV & 20.98 & 19.08 & 17.89 & 0.990 & 0.5138 & BOSS\\ 
DESI-037.0378-12.8812 & DEV & 20.39 & 19.01 & 17.92 & 0.997\\ 
DESI-038.2078-03.3906 & DEV & 20.08 & 18.39 & 17.40 & 0.999\\ 
DESI-040.3034-00.9646 & DEV & 19.27 & 17.83 & 17.02 & 0.999 & 0.2418 & SDSS\\ 
DESI-040.3168-06.5372 & DEV & 20.63 & 18.94 & 17.96 & 0.997\\ 
DESI-041.5205-06.1275 & DEV & 23.97 & 21.99 & 19.92 & 0.999\\ 
DESI-060.1043-16.3979 & DEV & 19.05 & 17.88 & 17.16 & 0.987\\ 
DESI-060.2420-13.9567 & DEV & 21.07 & 19.54 & 18.11 & 0.931\\ 
DESI-122.0852+10.5284 & DEV & 19.94 & 18.14 & 17.10 & 0.997 & 0.4754 & SDSS\\ 
DESI-133.5531-04.4026 & DEV & 21.80 & 20.45 & 18.72 & 0.849\\ 
DESI-133.6197+10.1374 & DEV & 19.06 & 17.44 & 16.57 & 0.989 & 0.2978 & SDSS\\ 
DESI-135.3125+09.9401 & COMP & 24.28 & 20.96 & 18.89 & 0.993\\ 
DESI-137.8568+14.2991 & DEV & 20.61 & 18.80 & 17.51 & 0.969 & 0.5464 & BOSS\\ 
DESI-140.8110+18.4954 & DEV & 20.20 & 19.60 & 18.65 & 0.875 & 0.8732 & BOSS\\ 
DESI-141.0626+05.7690 & DEV & 21.17 & 19.48 & 17.98 & 0.952 & 0.6128 & BOSS\\ 
DESI-143.3887+09.3219 & DEV & 22.16 & 20.41 & 18.71 & 0.999 & 0.7430 & BOSS\\ 
DESI-145.9507+00.9906 & COMP & 20.53 & 18.97 & 17.90 & 1.000\\ 
DESI-152.0763+31.7005 & COMP & 21.23 & 19.45 & 18.18 & 0.991 & 0.5394 & SDSS\\ 
DESI-154.6972-01.3590 & DEV & 20.53 & 18.72 & 17.79 & 1.000 & 0.3883 & BOSS\\ 
DESI-155.7366-05.0529 & DEV & 20.64 & 19.00 & 17.54 & 0.987\\ 
DESI-160.2351-01.0663 & DEV & 17.80 & 16.33 & 15.51 & 0.942 & 0.2502 & SDSS\\ 
DESI-162.4577+05.7749 & DEV & 18.76 & 17.37 & 16.55 & 0.997 & 0.2640 & SDSS\\ 
DESI-165.6876+12.1864 & DEV & 23.23 & 21.46 & 19.36 & 0.983\\ 
DESI-168.2943+23.9443 & DEV & 17.44 & 15.82 & 14.95 & 0.968 & 0.3361 & SDSS\\ 
DESI-186.3033-00.4390 & COMP & 19.49 & 18.17 & 17.44 & 0.987\\ 
DESI-186.8292+17.4324 & COMP & 21.56 & 19.95 & 19.01 & 0.903\\ 
DESI-189.5370+15.0309 & DEV & 20.38 & 18.69 & 17.37 & 0.998\\ 
DESI-194.5344+03.5358 & DEV & 19.90 & 18.29 & 17.31 & 0.989 & 0.4279 & BOSS\\ 
DESI-202.6690+04.6707 & DEV & 18.84 & 17.21 & 16.33 & 0.987 & 0.3363 & SDSS\\ 
DESI-204.0002-03.5250 & EXP & 21.36 & 21.12 & 20.37 & 0.900 & 0.1764 & SDSS\\ 
DESI-213.6664+19.4787 & DEV & 19.05 & 17.34 & 16.09 & 0.993 & 0.5768 & BOSS\\ 
DESI-214.6278+25.1814 & DEV & 17.83 & 16.23 & 15.41 & 0.986 & 0.2909 & SDSS\\ 
DESI-215.2654+00.3719 & DEV & 21.32 & 20.05 & 18.73 & 0.974\\ 
DESI-216.8280-06.7541 & DEV & 18.64 & 17.14 & 16.34 & 0.972\\ 
DESI-216.9538+08.1792 & DEV & 20.83 & 19.16 & 17.99 & 0.960 & 0.5338 & BOSS\\ 
DESI-219.7907+12.1404 & DEV & 20.54 & 18.60 & 17.59 & 0.863 & 0.4273 & SDSS\\ 
DESI-219.9855+32.8402 & DEV & 19.65 & 17.84 & 16.87 & 0.812 & 0.4176 & SDSS\\ 
DESI-318.0376-01.7568 & DEV & 18.14 & 16.73 & 15.92 & 0.974 & 0.2241 & BOSS\\ 
DESI-319.3483-00.9478 & DEV & 20.05 & 18.20 & 17.17 & 0.935 & 0.4272 & SDSS\\ 
DESI-349.5492-11.1012 & COMP & 19.51 & 17.78 & 16.12 & 0.980\\ 
DESI-359.8897+02.1399 & DEV & 18.94 & 17.04 & 16.05 & 0.993 & 0.4295 & BOSS\\ 
\enddata 
\tablecomments{\lensAz of the above \lensA Grade~A lens candidates have spectroscopic redshifts from SDSS (see text).  All redshift uncertainties $<3.7\times 10^{-4}$.} 
\end{deluxetable*}

%% file: table-grade-B.tex
\clearpage 
   \startlongtable 
   \begin{deluxetable*}{lccccccr} 
  \tabletypesize{\footnotesize} 
   \tablecaption{ 
   Grade B Candidates\label{tab:grade-B}} 
   \tablehead{ 
   \colhead{Name} & 
   \colhead{Type} & 
    \colhead{mag\_g} &  
   \colhead{mag\_r} &  
   \colhead{mag\_z} &  
   \colhead{Probability} & 
   \colhead{$z$} & 
   \colhead{Survey}} 
   \startdata 
DESI-005.6187+01.8037 & DEV & 20.93 & 20.09 & 18.59 & 0.889\\ 
DESI-009.1701+00.7687 & DEV & 21.91 & 20.44 & 18.82 & 0.964\\ 
DESI-009.9772-12.2100 & DEV & 21.93 & 20.06 & 18.91 & 0.900\\ 
DESI-010.2401-09.0954 & COMP & 21.70 & 20.14 & 18.99 & 0.990\\ 
DESI-010.2876-00.7303 & DEV & 21.33 & 19.67 & 18.40 & 0.898 & 0.5633 & SDSS\\ 
DESI-010.3630-01.1298 & DEV & 29.52 & 22.73 & 20.78 & 0.909\\ 
DESI-011.0219-04.8058 & DEV & 23.15 & 21.10 & 19.20 & 0.896 & 0.7715 & eBOSS\\ 
DESI-011.9235-06.1032 & DEV & 22.45 & 20.55 & 19.34 & 0.815\\ 
DESI-012.5310-18.6438 & DEV & 19.43 & 17.72 & 16.85 & 0.929\\ 
DESI-014.0105+03.0152 & DEV & 22.19 & 20.35 & 19.07 & 0.970 & 0.5367 & BOSS\\ 
DESI-014.0683-01.3924 & DEV & 20.16 & 18.96 & 17.75 & 0.989\\ 
DESI-014.0730-02.4262 & DEV & 20.98 & 19.34 & 18.41 & 0.994\\ 
DESI-014.4452-16.7457 & DEV & 21.64 & 19.79 & 18.56 & 1.000\\ 
DESI-014.6199-00.4627 & DEV & 21.53 & 20.40 & 19.21 & 0.951\\ 
DESI-015.0641-00.2718 & COMP & 21.00 & 19.31 & 18.33 & 1.000 & 0.4287 & SDSS\\ 
DESI-015.0721+00.0027 & DEV & 20.04 & 18.53 & 17.75 & 0.958 & 0.2758 & BOSS\\ 
DESI-015.2668-17.5344 & COMP & 20.63 & 19.31 & 18.41 & 0.997\\ 
DESI-015.2991+00.1673 & DEV & 21.55 & 19.97 & 18.99 & 0.985\\ 
DESI-015.3533-17.0192 & DEV & 22.03 & 20.28 & 19.11 & 0.902\\ 
DESI-015.3806+00.5700 & DEV & 22.56 & 20.98 & 19.40 & 0.987\\ 
DESI-015.4415+03.2399 & DEV & 20.93 & 19.42 & 18.39 & 0.968 & 0.5518 & BOSS\\ 
DESI-015.5430-00.3184 & DEV & 21.43 & 20.10 & 18.83 & 1.000 & 0.6376 & BOSS\\ 
DESI-015.6201-00.4207 & DEV & 20.71 & 19.09 & 18.21 & 0.998\\ 
DESI-015.6531-00.0963 & COMP & 20.75 & 19.11 & 18.21 & 1.000 & 0.3675 & BOSS\\ 
DESI-015.9831+00.3187 & DEV & 21.19 & 19.79 & 18.50 & 1.000 & 0.6205 & BOSS\\ 
DESI-016.1439-00.5185 & COMP & 18.66 & 17.01 & 16.14 & 1.000 & 0.3453 & SDSS\\ 
DESI-016.2119+00.4207 & DEV & 21.19 & 19.64 & 18.54 & 0.998 & 0.5293 & BOSS\\ 
DESI-016.2273+00.0668 & DEV & 19.58 & 18.07 & 17.19 & 0.999 & 0.2757 & SDSS\\ 
DESI-016.2921-18.3895 & DEV & 22.46 & 20.76 & 19.31 & 0.972\\ 
DESI-016.3969+00.1169 & COMP & 18.93 & 17.53 & 16.76 & 0.990\\ 
DESI-016.5355-00.2139 & DEV & 18.28 & 17.16 & 16.42 & 0.922 & 0.1971 & SDSS\\ 
DESI-016.7876+01.2914 & DEV & 20.25 & 18.39 & 17.41 & 0.999 & 0.4217 & BOSS\\ 
DESI-016.8597+03.2136 & COMP & 17.84 & 16.25 & 15.35 & 0.977 & 0.3245 & BOSS\\ 
DESI-016.9695-14.4480 & DEV & 21.30 & 19.38 & 18.07 & 0.999\\ 
DESI-017.0297-03.5796 & DEV & 19.77 & 18.34 & 17.54 & 0.985\\ 
DESI-017.4350-14.6600 & DEV & 21.17 & 19.54 & 18.02 & 0.995\\ 
DESI-017.5240-02.5417 & COMP & 19.31 & 17.49 & 16.53 & 0.933 & 0.4312 & BOSS\\ 
DESI-018.0754-04.5830 & COMP & 21.38 & 19.78 & 18.64 & 0.993 & 0.5290 & BOSS\\ 
DESI-018.1714-19.0457 & DEV & 20.58 & 18.93 & 17.49 & 0.999\\ 
DESI-018.2548-03.7210 & DEV & 20.00 & 18.34 & 17.45 & 0.992 & 0.3156 & BOSS\\ 
DESI-018.4039-18.9942 & DEV & 21.48 & 19.86 & 18.80 & 0.963\\ 
DESI-020.4712-17.9274 & COMP & 19.49 & 17.72 & 16.26 & 0.925\\ 
DESI-020.7598-13.2227 & DEV & 22.60 & 20.83 & 19.25 & 0.936\\ 
DESI-022.3389+00.6547 & DEV & 22.85 & 21.29 & 19.64 & 0.992\\ 
DESI-023.6765+04.5639 & DEV & 20.93 & 19.03 & 17.70 & 0.951 & 0.5508 & BOSS\\ 
DESI-029.0400-10.4926 & COMP & 22.79 & 21.24 & 19.85 & 0.999\\ 
DESI-034.9916-14.9460 & DEV & 22.88 & 21.30 & 19.64 & 0.927\\ 
DESI-035.0816-04.1971 & DEV & 19.83 & 18.39 & 17.63 & 0.994\\ 
DESI-035.7821-05.4661 & DEV & 22.21 & 20.37 & 19.23 & 0.999 & 0.4963 & BOSS\\ 
DESI-036.0677-16.3767 & DEV & 21.45 & 19.75 & 18.36 & 0.987\\ 
DESI-036.2542-05.6058 & DEV & 21.13 & 19.29 & 18.34 & 0.978 & 0.4381 & BOSS\\ 
DESI-036.3915-05.0365 & DEV & 19.25 & 18.14 & 17.43 & 0.998\\ 
DESI-036.4490-15.0922 & DEV & 21.60 & 20.11 & 18.71 & 0.959\\ 
DESI-037.0236-05.2927 & COMP & 20.58 & 19.48 & 18.77 & 0.999\\ 
DESI-038.9951-06.0696 & DEV & 22.83 & 20.96 & 19.60 & 0.988\\ 
DESI-040.5720-16.4116 & DEV & 21.02 & 19.15 & 17.76 & 0.939\\ 
DESI-040.7053-00.5888 & DEV & 22.08 & 20.22 & 19.24 & 1.000 & 0.4119 & BOSS\\ 
DESI-040.8111-00.1499 & DEV & 22.19 & 20.21 & 18.50 & 1.000 & 0.7167 & BOSS\\ 
DESI-041.4742-00.7052 & DEV & 21.23 & 20.09 & 19.39 & 1.000\\ 
DESI-041.9391-00.5247 & DEV & 21.25 & 19.38 & 18.06 & 1.000 & 0.5801 & SDSS\\ 
DESI-046.4723-14.8812 & DEV & 23.96 & 21.67 & 19.60 & 0.946\\ 
DESI-047.7087-17.7748 & DEV & 22.51 & 20.93 & 19.45 & 0.950\\ 
DESI-047.7647-13.2341 & DEV & 19.33 & 18.58 & 17.59 & 0.957\\ 
DESI-060.0471-15.8799 & DEV & 22.22 & 21.54 & 19.85 & 0.971\\ 
DESI-060.8033-15.2161 & COMP & 22.54 & 20.82 & 19.33 & 0.993\\ 
DESI-060.8089-15.0458 & DEV & 22.06 & 20.65 & 19.41 & 0.993\\ 
DESI-061.1134-17.2082 & DEV & 22.61 & 20.68 & 18.72 & 0.999\\ 
DESI-061.1909-14.5760 & COMP & 18.53 & 16.70 & 15.76 & 0.921\\ 
DESI-063.6323-04.5427 & DEV & 20.36 & 18.74 & 17.87 & 0.990\\ 
DESI-131.3607+00.0361 & DEV & 24.15 & 21.79 & 19.70 & 0.800\\ 
DESI-134.0057-07.2488 & DEV & 19.55 & 17.84 & 16.87 & 0.990\\ 
DESI-135.9714+07.1954 & DEV & 25.90 & 23.38 & 20.87 & 0.900\\ 
DESI-140.8863+20.3278 & DEV & 20.77 & 19.04 & 18.06 & 0.873\\ 
DESI-143.0565-05.6041 & DEV & 21.45 & 19.61 & 18.51 & 0.866\\ 
DESI-144.1511+08.8633 & COMP & 20.53 & 18.79 & 17.25 & 0.987\\ 
DESI-144.4242+31.4659 & COMP & 19.16 & 17.56 & 16.27 & 0.949 & 0.5969 & BOSS\\ 
DESI-144.6321-04.2535 & COMP & 21.15 & 19.36 & 18.03 & 0.921\\ 
DESI-145.0099+05.4279 & DEV & 21.35 & 20.12 & 18.86 & 0.900\\ 
DESI-150.0945+00.0047 & COMP & 18.92 & 18.64 & 18.18 & 1.000\\ 
DESI-150.8860-02.9493 & DEV & 20.89 & 18.95 & 17.36 & 0.992 & 0.6817 & BOSS\\ 
DESI-154.5307-00.1368 & DEV & 20.38 & 18.59 & 17.68 & 0.943 & 0.3718 & SDSS\\ 
DESI-154.7654+17.0697 & COMP & 18.73 & 17.16 & 16.32 & 0.939 & 0.3013 & BOSS\\ 
DESI-155.4865+11.2037 & DEV & 19.83 & 18.62 & 17.82 & 0.995\\ 
DESI-157.9622+01.7544 & COMP & 20.34 & 18.78 & 17.77 & 0.989\\ 
DESI-158.7893-02.3037 & COMP & 22.13 & 20.39 & 18.82 & 0.996\\ 
DESI-167.8517+14.1473 & DEV & 17.82 & 16.44 & 15.65 & 0.904 & 0.2211 & SDSS\\ 
DESI-170.6983+25.2669 & DEV & 20.00 & 18.11 & 17.12 & 0.983 & 0.4310 & SDSS\\ 
DESI-192.0242-06.5158 & COMP & 17.68 & 16.12 & 15.26 & 0.983\\ 
DESI-194.5900+15.6322 & DEV & 22.41 & 20.50 & 18.86 & 0.977 & 0.6847 & BOSS\\ 
DESI-194.8376+11.6490 & COMP & 19.62 & 18.33 & 17.51 & 0.935\\ 
DESI-201.7783+02.2129 & DEV & 20.33 & 18.81 & 17.98 & 0.636\\ 
DESI-201.7841-02.2996 & COMP & 21.79 & 19.97 & 18.25 & 0.931 & 0.7441 & BOSS\\ 
DESI-202.3729+31.3290 & DEV & 21.30 & 19.64 & 18.67 & 0.989\\ 
DESI-204.1663-05.7814 & DEV & 20.06 & 18.31 & 17.37 & 0.991\\ 
DESI-204.6057+28.3294 & DEV & 20.92 & 19.32 & 18.02 & 1.000 & 0.5841 & BOSS\\ 
DESI-211.0927+02.7242 & DEV & 21.47 & 19.83 & 18.92 & 0.951\\ 
DESI-216.1003+25.2423 & DEV & 20.53 & 19.06 & 18.27 & 0.989 & 0.2325 & BOSS\\ 
DESI-217.1429-07.0963 & DEV & 20.02 & 18.48 & 17.62 & 0.938\\ 
DESI-217.4784+12.0433 & DEV & 21.54 & 19.98 & 18.74 & 0.942 & 0.5531 & BOSS\\ 
DESI-219.0374-01.3295 & DEV & 20.10 & 18.66 & 17.60 & 0.847\\ 
DESI-219.9228+00.5073 & DEV & 18.67 & 17.67 & 17.00 & 0.837 & 0.1377 & SDSS\\ 
DESI-241.0592+06.4200 & DEV & 20.93 & 19.70 & 18.93 & 0.999\\ 
DESI-241.5432+14.1008 & COMP & 19.99 & 18.38 & 17.43 & 0.989\\ 
DESI-317.3884+05.1456 & COMP & 21.71 & 19.56 & 18.10 & 0.976 & 0.5642 & BOSS\\ 
DESI-328.5453+00.6329 & COMP & 18.81 & 17.62 & 16.92 & 0.998\\ 
DESI-351.4891-00.8741 & DEV & 23.36 & 21.27 & 19.52 & 0.982\\ 
\enddata 
\tablecomments{\lensBz of the above \lensB Grade~B lens candidates have spectroscopic redshifts from SDSS (see text).  All redshift uncertainties $<3.7\times 10^{-4}$.} 
\end{deluxetable*}

%% file: table-grade-C.tex
\clearpage 
   \startlongtable 
   \begin{deluxetable*}{lccccccr} 
  \tabletypesize{\footnotesize} 
   \tablecaption{ 
   Grade C Candidates\label{tab:grade-C}} 
   \tablehead{ 
   \colhead{Name} & 
   \colhead{Type} & 
    \colhead{mag\_g} &  
   \colhead{mag\_r} &  
   \colhead{mag\_z} &  
   \colhead{Probability} & 
   \colhead{$z$} & 
   \colhead{Survey}} 
   \startdata 
DESI-005.7434+00.1667 & DEV & 21.28 & 20.92 & 20.37 & 0.835\\ 
DESI-009.9958+00.5677 & DEV & 21.03 & 19.43 & 18.25 & 0.989 & 0.5255 & SDSS\\ 
DESI-010.2439-02.0572 & DEV & 20.22 & 19.19 & 18.57 & 0.917\\ 
DESI-013.5957-05.7105 & DEV & 21.44 & 19.59 & 18.44 & 0.961 & 0.5036 & BOSS\\ 
DESI-013.9264-01.0692 & DEV & 20.54 & 19.17 & 18.29 & 0.982\\ 
DESI-013.9873-00.6495 & DEV & 21.16 & 19.61 & 18.38 & 0.973 & 0.5659 & SDSS\\ 
DESI-014.6980+00.2355 & DEV & 21.83 & 20.16 & 19.24 & 0.996\\ 
DESI-014.7160-00.3509 & COMP & 18.77 & 17.67 & 16.99 & 0.997 & 0.2399 & SDSS\\ 
DESI-015.0586+01.8480 & DEV & 20.54 & 18.81 & 17.88 & 0.934 & 0.4046 & BOSS\\ 
DESI-015.1785-18.8779 & DEV & 21.37 & 19.92 & 19.14 & 0.908\\ 
DESI-015.1912+03.7221 & COMP & 20.21 & 18.37 & 17.43 & 0.970 & 0.3979 & BOSS\\ 
DESI-015.2403-16.9542 & COMP & 19.98 & 18.72 & 17.92 & 0.922\\ 
DESI-015.2570-17.7561 & DEV & 21.28 & 19.40 & 18.18 & 0.991\\ 
DESI-015.3628-00.9079 & COMP & 21.16 & 19.28 & 18.18 & 0.901 & 0.4635 & SDSS\\ 
DESI-015.3792-03.3438 & COMP & 19.82 & 18.69 & 17.99 & 0.962\\ 
DESI-015.8070-17.7531 & DEV & 20.75 & 18.99 & 18.07 & 0.931\\ 
DESI-015.9669-18.0271 & DEV & 21.10 & 19.46 & 18.53 & 0.987\\ 
DESI-016.1507-00.5780 & DEV & 21.26 & 19.72 & 18.79 & 0.990\\ 
DESI-016.1763-02.4491 & COMP & 22.77 & 20.87 & 18.88 & 0.936\\ 
DESI-016.1889-00.5809 & DEV & 21.79 & 20.10 & 19.26 & 0.990\\ 
DESI-016.2032-00.6226 & DEV & 21.65 & 19.85 & 18.86 & 0.973\\ 
DESI-016.2995-00.0690 & DEV & 21.93 & 20.32 & 19.27 & 0.994\\ 
DESI-016.3200+00.9163 & DEV & 21.36 & 19.78 & 18.85 & 0.923\\ 
DESI-016.4625+02.8614 & DEV & 21.49 & 19.84 & 18.87 & 0.913\\ 
DESI-016.4883-00.1774 & COMP & 22.85 & 21.02 & 19.35 & 0.992\\ 
DESI-016.5355-00.2139 & DEV & 18.28 & 17.16 & 16.42 & 0.922 & 0.1971 & SDSS\\ 
DESI-016.8009-16.8580 & DEV & 20.94 & 19.41 & 18.53 & 0.901\\ 
DESI-016.8554-00.7320 & COMP & 20.47 & 19.00 & 17.61 & 0.976\\ 
DESI-016.9199-05.3919 & DEV & 21.30 & 19.64 & 18.44 & 0.970 & 0.5214 & BOSS\\ 
DESI-016.9228-03.8680 & COMP & 21.06 & 19.46 & 18.18 & 0.979\\ 
DESI-016.9682+00.3567 & COMP & 19.59 & 17.97 & 17.09 & 1.000 & 0.3133 & SDSS\\ 
DESI-017.2362-17.9241 & DEV & 21.44 & 19.79 & 19.01 & 0.924\\ 
DESI-018.0415+00.1861 & DEV & 19.67 & 18.29 & 17.50 & 0.916\\ 
DESI-018.4701+03.4968 & DEV & 22.37 & 20.52 & 19.23 & 0.996 & 0.5517 & BOSS\\ 
DESI-019.8976-12.8253 & DEV & 21.96 & 20.12 & 18.95 & 0.945\\ 
DESI-020.3517+00.1654 & DEV & 21.01 & 19.92 & 18.75 & 0.971 & 0.7690 & BOSS\\ 
DESI-020.5273-04.8117 & DEV & 21.20 & 19.36 & 18.40 & 0.985 & 0.3978 & BOSS\\ 
DESI-021.5439-00.5883 & DEV & 21.33 & 19.79 & 18.95 & 0.985\\ 
DESI-021.8518-05.7560 & DEV & 21.75 & 20.15 & 19.11 & 0.944\\ 
DESI-021.9637+00.0975 & DEV & 21.43 & 20.15 & 19.39 & 0.967\\ 
DESI-022.6104-15.3370 & DEV & 21.47 & 19.70 & 18.37 & 0.980\\ 
DESI-023.6659-06.9344 & COMP & 22.22 & 20.65 & 19.76 & 0.921\\ 
DESI-023.7790+01.5306 & DEV & 22.82 & 21.13 & 19.60 & 0.921\\ 
DESI-024.7012-07.5349 & DEV & 21.38 & 19.67 & 18.61 & 0.975 & 0.4924 & BOSS\\ 
DESI-025.8361-10.0680 & COMP & 19.79 & 18.67 & 17.97 & 0.980 & 0.2422 & SDSS\\ 
DESI-026.2451-01.5934 & DEV & 22.38 & 20.80 & 19.62 & 0.905\\ 
DESI-026.2771-00.7284 & COMP & 20.29 & 19.39 & 17.97 & 0.949\\ 
DESI-027.5388-12.3284 & DEV & 20.88 & 19.15 & 18.28 & 0.926\\ 
DESI-027.8872-08.5764 & COMP & 21.94 & 20.07 & 18.71 & 0.945 & 0.5735 & BOSS\\ 
DESI-028.2710-09.8289 & DEV & 22.48 & 20.66 & 19.27 & 0.972 & 0.5973 & BOSS\\ 
DESI-028.3093-00.4385 & DEV & 22.39 & 21.07 & 19.39 & 0.958\\ 
DESI-028.8348-13.9044 & COMP & 20.24 & 19.28 & 18.71 & 0.900\\ 
DESI-029.1778-10.1834 & DEV & 22.01 & 20.90 & 19.47 & 1.000\\ 
DESI-030.8114-09.1587 & DEV & 20.11 & 18.62 & 17.87 & 1.000\\ 
DESI-031.8778-14.8046 & DEV & 24.15 & 21.92 & 20.34 & 0.954\\ 
DESI-033.9735-12.6841 & DEV & 21.85 & 20.04 & 18.87 & 0.992\\ 
DESI-034.0253-04.5771 & DEV & 22.25 & 20.46 & 19.29 & 1.000\\ 
DESI-034.3281-05.1331 & DEV & 23.33 & 21.65 & 20.26 & 0.993\\ 
DESI-035.7202-03.9575 & COMP & 22.69 & 21.50 & 19.82 & 1.000 & 0.8368 & eBOSS\\ 
DESI-035.8285-04.3979 & DEV & 21.41 & 19.72 & 18.80 & 0.937\\ 
DESI-035.8379-07.3794 & COMP & 19.89 & 17.97 & 16.63 & 0.995\\ 
DESI-035.8438-06.3246 & COMP & 20.05 & 18.32 & 17.47 & 0.967 & 0.3560 & BOSS\\ 
DESI-035.8660-18.4836 & COMP & 18.51 & 16.97 & 16.16 & 0.954\\ 
DESI-035.9027-06.8806 & COMP & 19.97 & 18.58 & 17.43 & 0.999\\ 
DESI-035.9185-05.8453 & DEV & 22.43 & 21.00 & 19.88 & 0.996\\ 
DESI-035.9393-04.4005 & DEV & 20.12 & 18.53 & 17.62 & 1.000 & 0.3037 & BOSS\\ 
DESI-036.0184-04.4084 & DEV & 21.71 & 20.19 & 19.39 & 1.000\\ 
DESI-036.0256-06.4796 & DEV & 21.58 & 20.29 & 19.47 & 0.986\\ 
DESI-036.0536-06.1645 & DEV & 20.81 & 19.19 & 18.34 & 0.979\\ 
DESI-036.0653-05.8029 & DEV & 22.90 & 21.55 & 19.62 & 0.998\\ 
DESI-036.0879+00.0726 & COMP & 19.59 & 18.29 & 17.52 & 0.959\\ 
DESI-036.1151-05.2254 & DEV & 22.06 & 21.00 & 19.61 & 0.998 & 0.9028 & eBOSS\\ 
DESI-036.1457-05.4990 & DEV & 21.98 & 20.87 & 19.50 & 0.991\\ 
DESI-036.2194-04.3486 & DEV & 22.68 & 21.65 & 19.95 & 1.000\\ 
DESI-036.2244-06.3029 & DEV & 22.39 & 20.74 & 19.54 & 0.956\\ 
DESI-036.2714-06.8192 & DEV & 19.62 & 18.92 & 18.38 & 0.983\\ 
DESI-036.3530-04.6792 & DEV & 19.01 & 17.48 & 16.67 & 1.000 & 0.2643 & BOSS\\ 
DESI-036.3795-04.2093 & DEV & 22.97 & 21.32 & 19.65 & 1.000 & 0.7726 & eBOSS\\ 
DESI-036.4031-04.2550 & DEV & 22.08 & 20.32 & 19.05 & 1.000 & 0.5557 & BOSS\\ 
DESI-036.4081-05.2473 & DEV & 20.63 & 19.04 & 18.17 & 0.999\\ 
DESI-036.4268-05.1548 & DEV & 19.80 & 18.63 & 17.94 & 1.000\\ 
DESI-036.4827-16.8621 & DEV & 19.41 & 17.99 & 17.14 & 0.984\\ 
DESI-036.6282-04.6316 & DEV & 22.26 & 20.51 & 19.22 & 1.000 & 0.5906 & BOSS\\ 
DESI-036.6556-03.7101 & DEV & 22.13 & 20.56 & 19.43 & 0.999\\ 
DESI-036.6760-03.6801 & DEV & 20.33 & 19.05 & 18.21 & 1.000\\ 
DESI-036.6777-03.6555 & COMP & 21.80 & 20.40 & 19.82 & 1.000\\ 
DESI-036.6819-03.6905 & DEV & 21.02 & 19.39 & 18.53 & 1.000 & 0.3284 & eBOSS\\ 
DESI-036.7198+00.2833 & DEV & 19.91 & 18.32 & 17.33 & 0.996 & 0.3022 & SDSS\\ 
DESI-036.7325-05.1276 & DEV & 21.73 & 19.87 & 18.76 & 1.000 & 0.4361 & BOSS\\ 
DESI-036.8076-05.0255 & DEV & 22.14 & 20.88 & 19.90 & 0.999\\ 
DESI-036.8133-03.9033 & COMP & 19.64 & 18.88 & 18.15 & 0.999\\ 
DESI-036.9560-06.1539 & DEV & 21.46 & 19.67 & 18.69 & 0.932 & 0.4324 & BOSS\\ 
DESI-037.0345-04.5015 & DEV & 22.20 & 20.50 & 19.45 & 1.000\\ 
DESI-037.1686-04.0027 & DEV & 20.09 & 18.75 & 18.01 & 1.000\\ 
DESI-037.2011-04.1161 & DEV & 20.17 & 19.28 & 18.67 & 0.993 & 0.1406 & BOSS\\ 
DESI-037.2064-01.2158 & DEV & 21.05 & 19.64 & 18.34 & 0.984 & 0.6895 & BOSS\\ 
DESI-037.2945+03.7522 & DEV & 24.56 & 22.05 & 19.98 & 0.945\\ 
DESI-037.4766+03.1089 & DEV & 22.26 & 20.84 & 19.46 & 0.985\\ 
DESI-038.2709-10.4498 & DEV & 22.03 & 20.38 & 19.19 & 0.957\\ 
DESI-038.8461-10.5125 & DEV & 20.76 & 19.29 & 18.51 & 0.964\\ 
DESI-039.0463-06.3428 & DEV & 22.71 & 21.00 & 19.42 & 0.973\\ 
DESI-039.2003+03.3583 & DEV & 20.93 & 19.31 & 18.25 & 0.984 & 0.4673 & BOSS\\ 
DESI-039.9261-01.4632 & DEV & 19.68 & 18.51 & 17.68 & 0.957\\ 
DESI-040.6372-12.1891 & DEV & 19.92 & 17.98 & 17.00 & 0.999\\ 
DESI-040.6769-00.6487 & DEV & 23.64 & 21.84 & 19.95 & 0.968\\ 
DESI-040.7046+02.2423 & DEV & 19.69 & 18.72 & 18.07 & 0.942\\ 
DESI-040.8745-01.9373 & DEV & 23.10 & 21.18 & 19.68 & 1.000\\ 
DESI-041.3678-01.2016 & DEV & 21.65 & 20.27 & 18.97 & 1.000 & 0.6669 & BOSS\\ 
DESI-041.4318-08.6492 & COMP & 21.44 & 20.05 & 18.64 & 0.999 & 0.7261 & BOSS\\ 
DESI-041.5548-00.7524 & COMP & 22.47 & 21.24 & 19.80 & 0.997\\ 
DESI-041.7915-08.4225 & DEV & 21.39 & 19.50 & 18.31 & 0.957 & 0.5221 & BOSS\\ 
DESI-041.9910-00.7425 & DEV & 21.70 & 20.45 & 19.68 & 0.999\\ 
DESI-042.2156-00.5329 & DEV & 19.33 & 18.02 & 17.22 & 0.995 & 0.2547 & BOSS\\ 
DESI-042.9152-00.5600 & DEV & 22.07 & 20.10 & 18.70 & 1.000 & 0.5846 & BOSS\\ 
DESI-046.6993-15.0593 & DEV & 22.53 & 20.97 & 19.64 & 0.985\\ 
DESI-047.3321-13.5368 & DEV & 22.19 & 20.49 & 19.21 & 0.998\\ 
DESI-060.4389-14.7568 & DEV & 20.04 & 18.57 & 17.77 & 0.986\\ 
DESI-060.6860-15.7303 & COMP & 19.67 & 18.89 & 18.30 & 0.976\\ 
DESI-061.0991-14.3883 & COMP & 22.79 & 21.03 & 18.97 & 0.954\\ 
DESI-064.4878-03.6133 & DEV & 19.04 & 17.83 & 17.07 & 0.997\\ 
DESI-125.6392-00.4650 & DEV & 21.11 & 19.20 & 17.94 & 0.951 & 0.5253 & BOSS\\ 
DESI-131.8556+14.2550 & COMP & 19.98 & 18.67 & 17.89 & 0.841\\ 
DESI-138.6664-00.0821 & 22.79 & 21.03 & 19.69 & 0.953\\ 
DESI-149.1942+00.7137 & DEV & 21.93 & 20.48 & 19.35 & 0.998\\ 
DESI-150.2022+01.6538 & DEV & 18.52 & 17.20 & 16.45 & 0.998\\ 
DESI-150.4045+02.5544 & DEV & 19.56 & 18.11 & 17.32 & 0.991 & 0.2477 & BOSS\\ 
DESI-151.2006-03.7158 & DEV & 20.00 & 18.64 & 17.92 & 0.911\\ 
DESI-151.7664+02.1430 & DEV & 20.51 & 18.84 & 17.95 & 0.972\\ 
DESI-151.9855+02.4052 & DEV & 21.90 & 20.17 & 18.97 & 0.943 & 0.5307 & BOSS\\ 
DESI-152.5264-01.9658 & DEV & 22.31 & 20.50 & 19.43 & 0.993\\ 
DESI-152.8042-02.0432 & DEV & 17.72 & 16.39 & 15.62 & 0.965\\ 
DESI-153.0462-00.8142 & COMP & 20.91 & 20.00 & 18.72 & 0.983\\ 
DESI-154.3116+02.4885 & DEV & 19.42 & 17.68 & 16.73 & 0.999 & 0.3576 & SDSS\\ 
DESI-155.4226+00.6966 & DEV & 22.48 & 20.58 & 19.12 & 0.919 & 0.6186 & BOSS\\ 
DESI-158.0944+15.8846 & DEV & 21.57 & 20.03 & 18.45 & 0.949\\ 
DESI-158.8311-00.5674 & DEV & 19.48 & 17.85 & 16.94 & 0.992 & 0.3157 & SDSS\\ 
DESI-170.8533+15.1850 & DEV & 19.04 & 17.36 & 16.46 & 0.987 & 0.3406 & BOSS\\ 
DESI-176.2181+08.9457 & DEV & 21.36 & 19.55 & 18.44 & 0.943 & 0.4971 & BOSS\\ 
DESI-180.0490-00.4182 & COMP & 23.41 & 21.54 & 19.96 & 0.971\\ 
DESI-181.9442+27.6152 & COMP & 18.56 & 17.04 & 16.20 & 0.995 & 0.3282 & SDSS\\ 
DESI-184.3703+15.6730 & DEV & 21.69 & 20.21 & 19.05 & 0.942\\ 
DESI-193.6112-08.7744 & DEV & 20.42 & 18.85 & 17.94 & 0.991\\ 
DESI-201.4063+04.1883 & DEV & 20.87 & 19.50 & 18.65 & 0.455\\ 
DESI-203.3751-02.1804 & DEV & 21.52 & 20.00 & 18.90 & 0.947\\ 
DESI-204.7174-08.3381 & DEV & 20.36 & 18.83 & 18.03 & 0.743\\ 
DESI-205.7370+22.6135 & DEV & 21.04 & 19.69 & 18.63 & 0.813 & 0.5198 & BOSS\\ 
DESI-210.3880+13.3370 & DEV & 20.12 & 18.83 & 18.05 & 0.856\\ 
DESI-212.6868-07.1025 & DEV & 20.45 & 19.12 & 18.34 & 0.942\\ 
DESI-240.0759+05.6966 & DEV & 21.18 & 19.44 & 18.52 & 0.995\\ 
DESI-240.3397+05.0773 & DEV & 21.87 & 20.91 & 19.89 & 0.981\\ 
DESI-240.4006+05.5796 & DEV & 20.43 & 18.97 & 18.21 & 0.993\\ 
DESI-240.5350+06.0657 & DEV & 22.56 & 20.99 & 19.66 & 1.000\\ 
DESI-240.7203+06.5371 & COMP & 21.62 & 19.97 & 18.67 & 0.997\\ 
DESI-241.2494+06.8555 & DEV & 21.68 & 20.27 & 19.36 & 0.999\\ 
DESI-241.3833+15.8226 & COMP & 19.76 & 18.27 & 17.22 & 0.969 & 0.5119 & BOSS\\ 
DESI-241.7841+07.0210 & DEV & 20.76 & 20.04 & 19.41 & 0.969\\ 
DESI-241.8463+07.1753 & COMP & 21.47 & 19.49 & 18.04 & 1.000 & 0.5903 & BOSS\\ 
DESI-242.0285+03.8786 & DEV & 22.07 & 21.07 & 19.44 & 0.909\\ 
DESI-242.4262+06.1599 & DEV & 22.07 & 20.20 & 18.97 & 0.996 & 0.5453 & BOSS\\ 
DESI-249.9825+19.0354 & COMP & 20.72 & 18.84 & 17.40 & 0.921 & 0.6064 & BOSS\\ 
DESI-251.0765+01.6752 & DEV & 21.32 & 19.46 & 18.41 & 0.921\\ 
DESI-251.1722+04.9724 & DEV & 20.45 & 18.90 & 17.99 & 0.901\\ 
DESI-317.2431+03.9841 & COMP & 19.70 & 18.96 & 18.34 & 0.957\\ 
DESI-319.7989+00.0575 & DEV & 20.16 & 18.79 & 18.00 & 0.958\\ 
DESI-338.0990+01.5111 & DEV & 22.37 & 20.79 & 19.51 & 0.992\\ 
DESI-351.1264-11.6503 & DEV & 21.83 & 19.94 & 18.55 & 1.000\\ 
DESI-351.1287-11.2566 & COMP & 21.80 & 20.75 & 19.72 & 0.998\\ 
DESI-351.1413-12.4955 & DEV & 20.75 & 18.92 & 17.98 & 1.000\\ 
DESI-351.2285-11.6281 & DEV & 21.89 & 20.09 & 19.11 & 0.999\\ 
DESI-351.2576-12.7728 & DEV & 24.17 & 22.08 & 19.89 & 0.997\\ 
DESI-351.3096-12.5492 & COMP & 20.71 & 19.13 & 18.29 & 1.000\\ 
DESI-351.3891-12.0013 & DEV & 20.74 & 19.38 & 18.59 & 0.998\\ 
DESI-351.4008-11.9943 & DEV & 21.19 & 19.36 & 18.44 & 0.998\\ 
DESI-351.4290-12.2431 & DEV & 20.59 & 18.78 & 17.88 & 1.000\\ 
DESI-351.4915-11.6013 & DEV & 20.47 & 19.17 & 18.41 & 0.998\\ 
DESI-351.5372-11.3464 & DEV & 21.76 & 20.16 & 19.38 & 0.999\\ 
\enddata 
\tablecomments{\lensCz of the above \lensC Grade~C lens candidates have spectroscopic redshifts from SDSS (see text).
All redshift uncertainties $< \zerrmaxC \times 10^{-4}$.} 
\end{deluxetable*}

%% file: discussion.tex
Our results so far are encouraging.  
\rf{In our current training sample we have only used 423 lenses.  
This is generally considered too small a number for a deep neural net.  
Nevertheless, we have succeeded in finding hundreds of new lens candidates over a large area of the sky for cutout images centered on elliptical galaxies.}
Here we will identify where we can improve.

For DR7, we have inspected the ResNet recommendations for approximately 3/4 of the sky for the DEV and COMP objects. 
For the remaining 1/4 of the folders, the number of recommendations are all high ($\gtrsim 1000$/folder), 
which, \rf{as mentioned in \S~\ref{sec:purity}, typically translates to 
low human inspection efficiency.}
We have stopped human inspection for now.  

\begin{minipage}{\linewidth}
\makebox[\linewidth]{
  \includegraphics[keepaspectratio=true,scale=0.24]{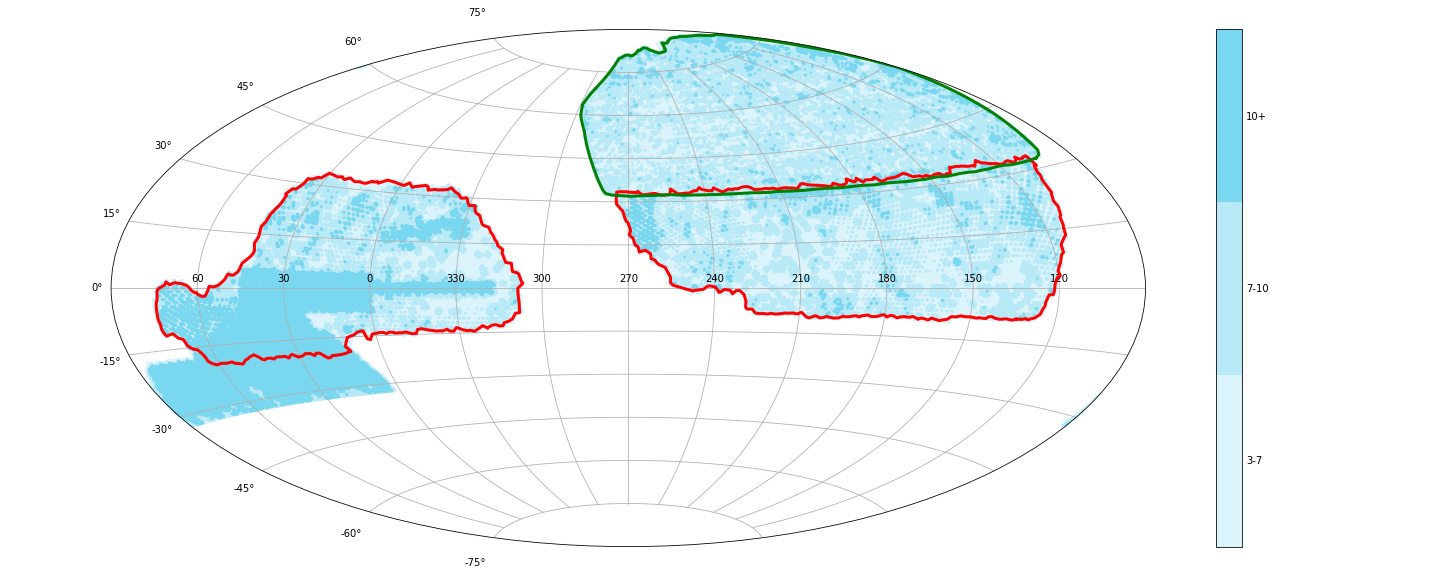}}
  \captionof{figure}{\rf{
  The coverage and depth map for DR8 of the Legacy Surveys.
  Below the Galactic plane, the footprint extends further south than DR7 (red outline), all of which overlaps with DES (also see Figure~\ref{fig:decals}).
  Note that this map includes the MzLS/BASS region.
  Color scheme for number of passes is the same as in Figure~\ref{fig:decals}.}}
  \label{fig:dr8-depth-map}
\end{minipage}

\rf{Based on our experience so far,
we believe we 
can build a much better training sample, retrain, and deploy
for DR8, which, as mentioned earlier, were completed during the referee procees for this paper.
The coverage and depth map for DR8 are shown in Figure~\ref{fig:dr8-depth-map}.
Note that there are patches where DR7 for DECaLS is deeper than DR8.
This is because it is desirable that DR8 has a more uniform coverage.}

\rf{\subsection{Toward a Better Training Sample}\label{sec:better-model}} 
\rf{We plan to take the following steps to improve our training sample.}
We need a larger 
non-lens sample (there are 13,000 in the current training sample) that will cover a greater variety of image configurations \rf{and,  
as mentioned in the previous section, include deep non-lens images.}
In our experience, having a large number of non-lenses is very helpful in terms of giving the neural net a better chance to reject a diverse variety of non-lenses.
This experience comports with what has been reported by other authors. 
\citet{metcalf2018a} used 20,000 simulated non-lenses in the Lens Competition, 
which do not include the many complications arising in real observations.
\citet{jacobs2019a} used 130,000 non-lenses in their training sample for an \rf{ensemble} of CNN models to find high redshift lenses in DES.
As stated in \S~\ref{sec:resnet}, 
so far we have not used the validation and testing sets in our training.
We have $\sim 600$ lenses in the entire current training sample.  
We can add \ed{$\sim 160$} lens candidates with high confidence, including all Grade~A and Grade~B candidates in this paper.  
Thus we will have $\sim 760$ lenses in our next training sample, $\sim 80\%$ more lenses than used for the current trained model.

In addition, we will make sure to include the following (many of these have been given greater than 0.9 probability of being a lens by the current trained ResNet model): star clusters; instances of interacting galaxies, ring galaxies, and more varieties of spiral galaxies than in the current training sample;
and more cosmic ray examples, especially those in the bluest ($g$) band and those with curved and/or thick tracks.




\rf{During the peer review process, we have switched our training to \rff{GPUs} and it is clear that}
the ResNet can train on a much larger sample within a reasonable amount of time.  
Though at this stage we have left the architecture and hyperparameters of the ResNet from L18 unchanged, 
we may vary both to optimize performance.

\rf{\subsection{The MzLS/BASS Region}}\label{sec:MzLS}
\rf{
The logical next step is to search for lensing systems in the northern MzLS/BASS region for the DEV and COMP categories.
In our current training sample, there are 21 lenses from the MzLS/BASS region, 
which have worse seeings in $gr$ bands.  
Here we briefly discuss the performance of our current model in this region.}
\rf{Out of a total of the eight lenses in the validation and testing sets (4 each) in MzLS/BASS, 
a total of three (1 in validation and 2 in testing) have probabilities $> 0.9$, or 37.5\%.  
This is remarkably similar to the 40\% completeness for DECaLS.  
Of course, this is an estimate based on small numbers.  
Nevertheless it provides a sense of the level of completeness for lenses discoverable in the MzLS/BASS region if we deploy our current trained model.
Efforts to adapt our model to this region, to account for the different seeing in $gr$ bands 
and therefore possibly improve the \rff{performance} of the ResNet model, is under way.}

\rf{\subsection{Lenses in the Tractor Galaxy Category of``REX"}\label{sec:REX}
The largest number of galaxies, by far, in the Legacy Surveys catalog are classified as the morphological type REX, i.e., the best-fit source model has a round exponential profile. 
The REX category contains an order of magnitude more objects than the DEV and COMP types combined, since most faint, extended galaxies are 
modeled by the REX profile.
It likely includes many elliptical galaxies,
though the percentage is unknown.  
Even so, given the total number of objects in this category is much larger than those in DEV and COMP categories, 
there will likely be a large number of lensing systems to be discovered.
\ed{Among the 199 known lenses from the Legacy Surveys in the training sample, 
18 are typed as REX.  Out of these, 12 have $z < 20.0$~mag and one with $z = 20.2$~mag.}
We will run inference on this category 
and report the results in a follow-up publication for lens search in DR8.
}


%% file: conclusion.tex

\ed{We have carried out a}
proof of concept end-to-end implementation of applying a deep residual neural network developed by \citet{lanusse2018a}, 
trained on observed lenses and non-lenses, to a subset of the Legacy Surveys data: 5.7 million elliptical galaxies from DECaLS with a $z$-band magnitude cut of 20.0 mag.
We use only real observations for training.  
In total, we have found \ed{\lensA Grade A candidates (of these, \lensAnew are new), \lensB Grade B and \lensC Grade C candidates (all new)}.



The results are promising. 
Despite using a relative small training set with 423 lens and 9451 non-lenses \rf{for a survey} with non-uniform depth (given that the survey was not yet completed),
in this paper we report the discovery of the first batch of \ed{\lenstotnew} new strong lens candidates from the Legacy Surveys.  
We will improve our training sample and model for the next round of training for full deployment on the 
entire Legacy Surveys DR8 
footprint.

%% file: acknowledgement.tex
We thank Steve Farrell, Mustafa Mustafa, Laurie Stephey, Rollin Thomas, and Prabat at the National Energy Scientific Computating Center (NERSC) for their consultation and advice. 
We thank Greg Aldering for insightful conversations in compiling our training sample.  
We thank Andi Gu for providing assistance for rendering some of the figures.   
We are grateful to Joel Brownstein and Lexi Moustakas for granting us access to the Master Lens Database (\url{http://admin.masterlens.org/index.php}).
This research used resources of the National Energy Research Scientific Computing Center (NERSC), a U.S. Department of Energy Office of Science User Facility operated under Contract No. DE-AC02-05CH11231 and the Computational HEP program in The Department of Energy's Science Office of High Energy Physics provided resources through the ``Cosmology Data Repository" project (Grant \#KA2401022).
X.H. acknowledges the University of San Francisco Faculty Development Fund. 
\ed{A.D.'s research is supported by the National Optical Astronomy Observatory, which is operated by the Association of Universities for Research in Astronomy (AURA) under cooperative agreement with the National Science Foundation.}

\ed{This paper is based on observations at Cerro Tololo Inter-American Observatory, National Optical
Astronomy Observatory (NOAO Prop. ID: 2014B-0404; co-PIs: D. J. Schlegel and A. Dey), which is operated by the Association of
Universities for Research in Astronomy (AURA) under a cooperative agreement with the
National Science Foundation.}

\ed{This project used data obtained with the Dark Energy Camera (DECam),
which was constructed by the Dark Energy Survey (DES) collaboration.
Funding for the DES Projects has been provided by 
the U.S. Department of Energy, 
the U.S. National Science Foundation, 
the Ministry of Science and Education of Spain, 
the Science and Technology Facilities Council of the United Kingdom, 
the Higher Education Funding Council for England, 
the National Center for Supercomputing Applications at the University of Illinois at Urbana-Champaign, 
the Kavli Institute of Cosmological Physics at the University of Chicago, 
the Center for Cosmology and Astro-Particle Physics at the Ohio State University, 
the Mitchell Institute for Fundamental Physics and Astronomy at Texas A\&M University, 
Financiadora de Estudos e Projetos, Funda{\c c}{\~a}o Carlos Chagas Filho de Amparo {\`a} Pesquisa do Estado do Rio de Janeiro, 
Conselho Nacional de Desenvolvimento Cient{\'i}fico e Tecnol{\'o}gico and the Minist{\'e}rio da Ci{\^e}ncia, Tecnologia e Inovac{\~a}o, 
the Deutsche Forschungsgemeinschaft, 
and the Collaborating Institutions in the Dark Energy Survey.
The Collaborating Institutions are 
Argonne National Laboratory, 
the University of California at Santa Cruz, 
the University of Cambridge, 
Centro de Investigaciones En{\'e}rgeticas, Medioambientales y Tecnol{\'o}gicas-Madrid, 
the University of Chicago, 
University College London, 
the DES-Brazil Consortium, 
the University of Edinburgh, 
the Eidgen{\"o}ssische Technische Hoch\-schule (ETH) Z{\"u}rich, 
Fermi National Accelerator Laboratory, 
the University of Illinois at Urbana-Champaign, 
the Institut de Ci{\`e}ncies de l'Espai (IEEC/CSIC), 
the Institut de F{\'i}sica d'Altes Energies, 
Lawrence Berkeley National Laboratory, 
the Ludwig-Maximilians Universit{\"a}t M{\"u}nchen and the associated Excellence Cluster Universe, 
the University of Michigan, 
{the} National Optical Astronomy Observatory, 
the University of Nottingham, 
the Ohio State University, 
the OzDES Membership Consortium
the University of Pennsylvania, 
the University of Portsmouth, 
SLAC National Accelerator Laboratory, 
Stanford University, 
the University of Sussex, 
and Texas A\&M University.}
